\documentstyle[12pt]{article}
\input{epsf}

\newcommand{\ba}{\begin{array}}
\newcommand{\ea}{\end{array}}
\newcommand{\as}{\alpha_s}

\newcommand{\msbar}{\overline{\mbox{MS}}}
\newcommand{\dsp}{\displaystyle}
\def\ltap{\raisebox{-.4ex}{\rlap{$\sim$}} \raisebox{.4ex}{$<$}}
\def\gtap{\raisebox{-.4ex}{\rlap{$\sim$}} \raisebox{.4ex}{$>$}}
\def\etal{{\it et al. \ }}
\newcommand{\apib}{\frac{\bar{\alpha}_s}{\pi}}

\begin{document}
\begin{titlepage}
\today          \hfill 
\begin{center}
\hfill    LBL-38549 \\

\vskip .5in

\noindent
{\large \bf Parton Model Sum Rules}\footnote{This work was
 supported by the Director, Office of Energy
Research, Office of High Energy and Nuclear Physics, Division of High
Energy Physics of the U.S. Department of Energy under Contract
DE-AC03-76SF00098.}

\vskip .5in

\vskip .5in

Ian Hinchliffe and Axel Kwiatkowski\\

{\em Theoretical Physics Group\\
    Ernest Orlando Lawrence Berkeley National Laboratory\\
      University of California\\
    Berkeley, California 94720}
\end{center}

\vskip .5in

\begin{abstract}
This  review article discusses the experimental and theoretical status of
various Parton Model sum rules. The basis of the sum rules in perturbative
 QCD
is discussed. Their use in extracting the value of the strong coupling 
constant
is evaluated and the failure of the naive version of some of these rules 
is 
assessed.
\end{abstract}
\end{titlepage}
\renewcommand{\thepage}{\roman{page}}
\setcounter{page}{2}
\mbox{ }

\vskip 1in

\begin{center}
{\bf Disclaimer}
\end{center}

\vskip .2in

\begin{scriptsize}
\begin{quotation}
This document was prepared as an account of work sponsored by the United
States Government. While this document is believed to contain correct 
 information, neither the United States Government nor any agency
thereof, nor The Regents of the University of California, nor any of their
employees, makes any warranty, express or implied, or assumes any legal
liability or responsibility for the accuracy, completeness, or usefulness
of any information, apparatus, product, or process disclosed, or represents
that its use would not infringe privately owned rights.  Reference herein
to any specific commercial products process, or service by its trade name,
trademark, manufacturer, or otherwise, does not necessarily constitute or
imply its endorsement, recommendation, or favoring by the United States
Government or any agency thereof, or The Regents of the University of
California.  The views and opinions of authors expressed herein do not
necessarily state or reflect those of the United States Government or any
agency thereof, or The Regents of the University of California.
\end{quotation}
\end{scriptsize}

\vskip 2in

\begin{center}
\begin{small}
{\it Lawrence Berkeley Laboratory is an equal opportunity employer.}
\end{small}
\end{center}

\newpage
\renewcommand{\thepage}{\arabic{page}}
\setcounter{page}{1}

\section{Introduction}
One of the best tools to use in attempting to disentangle the structure 
of the nucleon is lepton-nucleon scattering
where the lepton, whose couplings to electroweak gauge bosons is fully 
known,  is used as a probe on the constituents of the nucleon. 
Lepton-nucleon scattering with large momentum transfer between the
lepton and the nucleon is described 
in terms of the  Parton Model \cite{bjorken69,feynman69}. In its naive
 form this
model describes the nucleon as a collection
 of non-interacting quarks and gluons.
Lepton-nucleon scattering is then viewed
 as the sum of incoherent scatterings by the
lepton off these partonic constituents.  
 The description of these constituents is most conveniently given
in a frame where the nucleon has large momentum.
 If the nucleon mass is neglected, its momentum can be written as
 $P^\mu=(p,p,0,0)$.
A parton momentum can be written as
 $P^\mu_i=(z_ip,z_ip,p_t,0)$ where $p_t\sim 300$ MeV is related to the
scale of nucleon binding.  A distribution
 function $f_i(z)$ is defined so that the probability that
a parton of type $i$ (for example an up quark)
 has momentum in the range $P(z)$ to $P(z+dz)$
is $f_i(z)$. The lepton-nucleon scattering
 rates are then expressed in terms of $f_i$.

The target nucleon is characterised by certain 
quantum numbers such as isospin and baryon number. These quantum 
numbers are carried
by  the      constituents.  For    example,  the   net   number 
 of  up   quarks  in  a   proton  is   two,   hence    $\int_0^1 dz
(f_u(z)-f_{\overline{u}}(z))=2$. By forming appropriate  combinations 
of scattering cross-sections, quantities can be measured that
correspond to these conserved quantum numbers and hence have 
simple values in the Naive Parton Model. These quantities are referred
to as Parton Model sum rules. They can  then be compared with
 experiment and the  fundamental properties of theory tested. This
Naive  Parton  Model is  subject  to  corrections  in the  full 
 theory of  strong   interactions  (Quantum   Chromodynamics or QCD
\cite{politzer,gross-wil}, for recent reviews see
 \cite{Alt82,Rey81,Bur80} ). 
These corrections fall into two types; those that are
strongly  suppressed at  high energy  (higher twist   corrections)
 and those  that vanish  only  logarithmically  with the momentum
transfer. The latter are  fully calculable in terms  of the
 coupling constant  $\alpha_s$ of QCD.  Comparison of the sum rules with
these QCD expectations then provides a powerful test of QCD and
 enables $\alpha_s$ to be measured.

In the remainder of this article we will discuss these sum rules. We
show their 
values in the Naive Parton Model and the corrections from QCD and
finally compare
these predicted values with experiment.
\subsection{\it Deep Inelastic Scattering}
The kinematics of lepton-nucleon scattering are shown in Figure 
1.
The scattering of an unpolarized charged lepton (an electron or muon)
or neutrino
of momentum $k$ off an unpolarized
 nucleon of mass $M$, momentum $P$ results
in a final state
 with a lepton of momentum $k^{\prime}$ and a nuclear fragment.
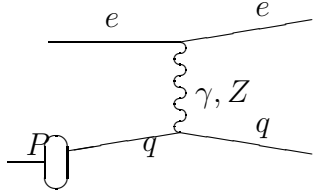
\begin{figure}
\setlength{\unitlength}{0.5mm}
\begin{picture}(120,70)
\put(25,54){\line(1,0){35}}
\put(60,54){\line(6,1){35}}
\multiput(60,31.5)(0,6){4}{\oval(3,3)[l]}
\multiput(60,34.5)(0,6){4}{\oval(3,3)[r]}
\put(60,30){\line(6,-1){35}}
\put(60,30){\line(-6,-1){30}}
\put(24,22){\line(-1,0){10}}
\put(30,21){\line(3,-1){7}}
\put(30,19){\line(3,-1){7}}
\put(27,22){\oval(6,14)}
\put(40,58){$e$}
\put(80,61){$e$}
\put(64,38){$\gamma,Z$}
\put(50,25){$q$}
\put(80,30){$q$}
\put(19,24){$P$}
\end{picture}
\label{dfd}
\caption[]{ Diagram illustrating the deep inelastic scattering process
{\mit electron + proton $\to$ electron + anything}.}
\end{figure}
The inclusive cross-section for this process can be described by two
Lorentz  scalars which can be taken to be $Q^2=-(k^{\prime}-k)^2$ and
$\nu=P\cdot
 (k-k^{\prime})$. Convenient dimensionless
variables are $x=Q^2/2\nu$ and $y=\nu/P\cdot k$ both of which
range from zero to one. The latter is 
the fractional energy loss of the lepton in the rest frame of the nucleon.
Two types of scattering are important, neutral and charged current scattering.
The former case describes the process $\mu N \to \mu X$ 
or $e N\to e X$. 
For neutral current
 scattering and $Q^2<<M_Z^2$ (all of the data discussed in this article
satisfy this requirement), the cross-section can be expressed as
\begin{equation}
\begin{array}{rl}
\dsp\frac{d\sigma^{eN}}{dxdy}&=\dsp \frac{4\pi\alpha_{em}^2}{sx^2y^2}\cr
\dsp (xy^2F_1^{eN}(x,Q^2)&+(1-y-M^2x^2y^2/Q^2)F_2^{eN}(x,Q^2))
\end{array}
\label{eN}
\end{equation}
where $s=(P+k)^2$ and $F_1,F_2$ are arbitrary functions
 called structure functions. This is the most general form that
this cross-section can take 
consistent with 
Lorentz invariance and parity conservation provided that
terms that are proportional 
to quark masses (actually $m_q^2/Q^2$) are neglected. This is an
excellent approximation since the nucleon consists
 mainly of up, down and strange quarks that have very 
small masses ($m_u\sim m_d\sim \hbox{few}$ MeV 
and $m_s\sim 100$ MeV \cite{quarkmass}).
For charged current scattering, the
 presence of parity violation which 
proceeds via the
exchange of a $W$ boson between the lepton and the target nucleon,  
requires the appearance of another function $F_3$.
\begin{equation}
\begin{array}{ll}
\dsp \frac{d\sigma^{\nu N}}{dxdy}=&\dsp\frac{G_F^2M_W^4s}
{2\pi (Q^2+M_W^2)^2}(xy^2F_1^{\nu N}(x,Q^2)\cr
&\dsp +(1-y-x^2  y^2M^2/Q^2)F_2^{\nu N}(x,Q^2)\cr
& \dsp -\frac{1}{2}x((1-y)^2-1)F_3^{\nu N}(x,Q^2))
\end{array}
\end{equation}
For $\overline{\nu}N$ scattering 
the sign of the last ($xF_3$) term is reversed.
The above formulae assume that the target nucleon and charged lepton
 are  unpolarized (the neutrino is always polarized).
In the case 
of $eN$ scattering with polarized electron and nucleon a parity
{\it conserving} asymmetry can be formed.
\begin{equation}
 a(x,y)=\frac{d\sigma^{e N}_p}{dxdy}-\frac{d\sigma^{e N}_{ap}}{dxdy}
\end{equation}
where the subscript $p$ ($ap$) refers to
 the state where the nucleon spin is parallel (anti-parallel) to 
its direction of motion in the center of mass frame 
of the lepton-nucleon  system. In both cases the lepton has
its spin aligned along its direction of motion. Then
\begin{equation}
\begin{array}{ll}
a(x,y)=&\dsp\frac{8\pi \alpha_{em}^2 y}{MQ^2}
((1-2/y^2+2x^2y^2M^2/Q^2)G_1(x,Q^2)\\
&+4x^2M^2G_2(x,Q^2)/Q^2)
\end{array}
\label{asym}
\end{equation}
defines the spin dependent structure functions $G_1$ and $G_2$.

In the rest frame of the target nucleon, 
the various kinematic quantities are related to the
energy $E$ ($E^{\prime}$) of the incoming
 (outgoing) lepton and the scattering angle $\theta$ of the lepton by
$y=\nu/ME=(E-E^{\prime})/E$ and $\sin^2(\theta/2)=Q^2/(4EE^{\prime})$.
By measuring the incoming 
and outgoing lepton energy and the scattering angle
the structure functions can be determined. 
\subsection{\it Quark Parton Model}
In the Quark Parton Model, lepton-nucleon scattering is described by
the scattering of 
 a lepton off the partonic ({\it i.e.} quark and gluon) constituents
of the nucleon. The nucleon structure is described in terms of 
the parton distribution functions $f_i(z)$.
We shall often use the following symbols to simplify the notation :-
$f_i(z)\equiv i(z)$ {\it etc.} so
 that the up quark distribution in a proton is $u(z)$
  and the anti-down
quark distribution is $\overline{d}(z)$. 
These distributions 
will always refer to a proton. When a neutron target is
involved we use isospin symmetry to relate the  neutron distributions to
those of the proton so that the up quark distribution in a neutron is
the down quark distribution in the proton and vice-versa, while
the gluon ($g(z)$) and the strange, charm, bottom and top distributions  
($s(z),c(z),b(z),t(z)$) are the same in proton and neutron. 
 The distribution of these heavier quarks is smaller and in the
case of top and to a lesser extent bottom totally negligible.
The charm quark is troublesome since its mass cannot be neglected in
experiments that have $Q^2\sim M_c^2\sim (1.5 \hbox{ GeV})^2.$
This effect is most important in neutrino scattering where the process
$\nu+s\to \mu^- + c$ is a significant part of the cross-section. 

Since the quarks carry the quantum numbers of the nucleon these
distribution functions satisfy certain constraints. For example,
the electric charge of the proton yields
\begin{equation}
1=\int_0^1 dz(\frac{2}{3}
(u(z)-\overline{u}(z))-\frac{1}{3}(d(z)-\overline{d}(z))
-\frac{1}{3}(s(z)-\overline{s}(z)))
\end{equation} 
and the zero net strangeness of the proton gives 
\begin{equation}
0=\int_0^1 dz(s(z)-\overline{s}(z))
\end{equation} 
Note that this does not imply that $s(z)=\overline{s}(z)$.
Momentum conservation in the lepton-parton scattering process implies that
the parton momentum fraction $z$ is identified 
with the kinematic variable $x$ in the Naive Parton Model.
The structure functions $F_1$, $F_2$ and $F_3$ are given in terms of the
parton distribution functions. In particular 
\begin{equation}
F_2(x,Q^2)^{ep}=x\sum_i q_i^2 f_i(x) 
\label{f2parton}
\end{equation} 
where $q_i$ is the charge of the type $i$ parton. The Naive Parton Model
therefore predicts
 that there is no $Q^2$ dependence in $F_i(x,Q^2)$. The partons
that couple
 to the photon and $W$ boson (quarks) have spin $1/2$ and hence
$F_2(x,Q^2)=2xF_1(x,Q^2)$, 
the Callan Gross relationship \cite{callan74}.

The relationships between
 the structure functions for neutrino scattering and the
parton distributions are complicated by
 the Kobayashi-Maskawa \cite{km-matrix} mixing matrix,
$V_{ij}$
which determines the
 relative strength of the  coupling of a quark pair to the $W$
boson :- $W^{\mu}\overline{q}_i\gamma_\mu 
(1-\gamma_5)q_j V_{ij}$. In order 
to simplify the results that follow, the equations
 are written
 in the approximation that the mixing matrix is a diagonal unit
 matrix. The mixing can be added by replacing the
 quark distributions by appropriate
linear combinations,
 for example $d(z) \to (V_{du}+V_{dc}+V_{dt} )d(z)$. 
 Neutrino scattering 
proceeds off up and anti- down type quarks {\it viz}
$\nu u\to \mu^+ d$ and 
$\nu \overline{d}\to \mu^+ \overline{u}$ leading to the
following relations (neglecting top and bottom quark contributions)
\begin{equation}
F_3^{\nu p}(x,Q^2)=2(d(x)-\overline{u}(x)-\overline{c}(x)+s(x)) 
\end{equation} 
and
\begin{equation}
F_2^{\nu p}(x,Q^2)=2x(\overline{u}(x)+\overline{c}(x)+d(x)+s(x))
\end{equation} 
The other relations can be trivially obtained from these.
 The charm quark contributions will not be written explicitly in the following.

The spin structure of the nucleon is probed in polarized scattering. One can
define $\Delta f_i(z)$ as the difference between the  parton distributions for 
parton of type $i$ with  spins (helicity) parallel and anti parallel to the
nucleon's spin. The unpolarized distributions introduced above are then the sum
of these two states. 
Define the quantities $a_0$, $a_3$ and $a_8$ by
\begin{equation}
\begin{array}{ll}
a_0&\dsp
=
\int_0^1 dx \Big(\Delta u + \Delta \bar{u}
                +\Delta d + \Delta \bar{d}
                +\Delta s + \Delta \bar{s} \Big)\cr
 
 \dsp
a_3
& \dsp
= \int_0^1 dx \Big(\Delta u + \Delta \bar{u}
                -\Delta d - \Delta \bar{d} \Big)
\\ \dsp
a_8
& \dsp
= \frac{1}{\sqrt{3}}
\int_0^1 dx \Big(\Delta u + \Delta \bar{u}
                +\Delta d + \Delta \bar{d}
                -2\Delta s -2 \Delta \bar{s} \Big)
\\ \dsp
\label{a0eq}
\end{array}
\label{aeqn}
\end{equation} 
These quantities are related to  the matrix elements of the axial vector current
between nucleon states. The matrix
 elements involving changes of flavor  can be determined from weak decays.
$a_3$ is determined from neutron $\beta$-decay:
$a_3 = g_A = \frac{G_V}{G_A}=1.2573\pm 0.0028$. 
$a_8$ and $a_3$  can be constrained
from the weak decay constants
  of ($\Sigma, \Lambda$ and $\Xi$) hyperons \cite{pdg}.
Assuming $SU(3)_F$ symmetry for the octet axial vector currents this 
gives $F=(\sqrt{3}a_8+a_3)/4=0.459\pm 0.008$ and 
$D=\sqrt{3}(\sqrt{3}a_3-a_8)/4=0.798\pm 0.008$
implying $a_8=0.33 \pm 0.02$ \cite{close93}.
Data from $\nu p$ and $\overline{\nu}p$ elastic scattering 
\cite{ahr87} provide information on the matrix element
of the flavor
 singlet current $a_0$ to be determined \cite{kapman88,ellkarl88}.
These data do not directly
 measure the static
 quantities and a form factor behaviour must be assumed \cite{alber95}.
In addition the experimental errors are large. The data can be interpreted as
$\int_0^1 dx \Big(\Delta s + \Delta \bar{s} \Big)=\Delta S=-0.15 
 \pm 0.09$ \cite{kapman88}
which (using the above value of $a_8$) implies $a_0=0.12 \pm 0.27$.
If all of the nucleon's spin 
is carried by quark 
spin and not by gluons or orbital angular momentum one expects that $a_0=1$.
In the Parton Model the spin structure functions are
\begin{equation}
\begin{array}{ll}
G_2(x,Q^2)&=0\cr 
G_1^{ep}(x,Q^2)=&\frac{1}{2}\sum_iq_i^2\Delta f_i(x) 
\end{array}
\end{equation} 
\subsection{\it Sum Rules}
In this section we list the various sum rules
and their values in the Naive Parton Model. These rules are all derived from
inclusive quantities that have a simple interpretation in this model.
We will refer to these sum rules by  a name that relates to this simple
interpretation. We will also
 indicate the more
 familiar names by which they are sometimes referred in the literature.

The {\sc baryon (Gross Llewellyn-Smith) sum rule}
 \cite{GLS69} uses the average of 
$F_3$ measured on a proton
 and a neutron. Since neutrino experiments are often
performed on heavy
 nuclear targets (such as iron) which have an almost equal
number of protons
 and neutrons, the quantity is  readily measured. The sum rule
measures the sum of
 the baryon number ($B$) and strangeness ($S$) of the nucleon.
\begin{equation}
\ba{ll}\dsp
S_{{\rm GLS}}
& \dsp
= \int_0^1 dx F_3^{\nu{\cal N}}
=\int_0^1 dx \frac{1}{2}\left(
              F_3^{\nu p}+F_3^{\nu n}
                       \right)\cr
&\dsp
=\int_0^1 dx \Big( u(x)+d(x)-\bar{u}(x)-\bar{d}(x)+2(s(x)-\bar{c}(x)) \Big)
\\ & \dsp
= 3
\ea
\end{equation}
The  {\sc Isospin  (Adler) sum rule}
 \cite{Adl66} measures the isospin of the target and
depends on the difference in $F_2$ measured in neutrino scattering off
proton and neutron targets.
\begin{equation}
\ba{ll}\dsp
S_{{\rm A}}
& \dsp
=  \int_0^1 \frac{dx}{x} \left(
        F_2^{\nu n}-F_2^{\nu p}
            \right)
\\ & \dsp
=2 \int_0^1 dx \Big( u(x)-d(x)-\bar{u}(x)+\bar{d}(x) \Big)
\\ & \dsp
= 4 I_3=2
\ea
\end{equation}
A similar sum rule can be formed in electron scattering.
We can introduce
``valence'' distributions 
defined by $u_v(z)\equiv u(z)-\overline{u}(z)$ and 
$d_v(z)\equiv d(z)-\overline{d}(z)$.
 Note that, since the net number of up (down)
quarks in a proton is 2 (1), 
these satisfy $\int dzu_v(z)=2$ and $\int dzd_v(z)
=1$.  Then
\begin{equation}
\ba{ll}\dsp
S_{{\rm G}}
& \dsp
= \int_0^1 \frac{dx}{x} \left(
        F_2^{ep}-F_2^{en}
            \right)
\\ & \dsp
= \frac{1}{3}
\int_0^1 dx \Big( u(x)-d(x)+\bar{u}(x)-\bar{d}(x) \Big)
\\ & \dsp
= \frac{1}{3}
\int_0^1 dx \Big( u_v(x)-d_v(x)\Big)
+ \frac{2}{3}
\int_0^1 dx \Big( \bar{u}(x)-\bar{d}(x) \Big)
\\ & \dsp
= \frac{1}{3}
\ea
\label{naive-got}
\end{equation}
The last step
 follows if $\int dz \overline{d}(z)=\int dz \overline{u}(z)$.
However, there is no fundamental reason for this simple assumption to
be valid and hence the 
{\sc Valence Isospin (Gottfried) sum rule} \cite{Got67} 
is on much weaker ground than the previous one.

The total momentum carried by all of the proton's constituents 
is constrained to add up to that of the proton. Hence the
{\sc Momentum} sum rule.
\begin{equation}
\begin{array}{ll}
S_{{\rm mom}} =& \int_0^1 dx\; x \;\Big(
 u(x)+d(x)+s(x)+\bar{u}(x)+\bar{d}(x)+\bar{s}(x)
 \Big)\cr 
&= 1-\int_0^1 dx\; x g(x)
\end{array}
\end{equation}
This sum rule 
cannot be tested directly since the gluon distribution function
does not appear
 in the structure functions. Rather, the left hand side of this
equation can be
 measured and the rule used to infer something about the gluon
distribution.

The {\sc Second Isospin (Unpolarized Bjorken) sum rule} \cite{Bjo67} is
\begin{equation}
\ba{ll}\dsp
S_{{\rm Bj}}^{{\rm unpol.}}
& \dsp
= \int_0^1 dx \left(
        F_1^{\nu n}-F_1^{\nu p}
            \right)
\\ & \dsp
=\int_0^1 dx \Big( u(x)-d(x)-\bar{u}(x)+\bar{d}(x) \Big)
\\ & \dsp
= 1
\ea
\end{equation}
It is equivalent to the {\sc Isospin sum rule} in the Naive Parton Model
where $F_2(x,Q^2)=2xF_1(x,Q^2)$. This equivalence is broken 
once QCD corrections
are computed.

The remaining sum
 rules depend on the spin structure function $G_1(x,Q^2)$ and 
therefore on the
 polarized distribution functions. The integrals  are related
to
the quantities $a_0$, $a_3$ and $
a_8$ by  
\begin{equation}
\int_0^1 dx G_1^{p(n)} =
\frac{1}{12}\left(
   \pm a_3 + \frac{1}{\sqrt{3}}a_8 + \frac{4}{3}a_0
            \right)
\end{equation}

The {\sc Polarized Isospin  (Bjorken Spin) sum rule }\cite{Bjo66}:
\begin{equation}
S_{{\rm Bj}}
=\int_0^1 dx \left( G_1^{p}-G_1^{n}\right)
=\frac{1}{6}a_3
\end{equation}
If we assume that the strange quarks are unpolarized  
$\Delta s = 0 \rightarrow a_0 = \sqrt{3}a_8$
then we have the {\sc Spin (Ellis Jaffe) sum rules} \cite{Gou72,EllJaf74}
\begin{equation}
\ba{lll}\dsp
S_{{\rm EJ}}^p
& \dsp
=\int_0^1 dx  G_1^{p}
& \dsp
= \frac{1}{12}\left(
   a_3+\frac{5}{\sqrt{3}}a_8
                     \right)
\\ \dsp
S_{{\rm EJ}}^n
& \dsp
= \int_0^1 dx  G_1^{n}
& \dsp = \frac{1}{12}\left(
   -a_3+\frac{5}{\sqrt{3}}a_8
                     \right)
\ea
\end{equation}
The Parton Model prediction $G_2(x,Q^2)=0$ leads to a trivial prediction
for the 
{\sc $G_2$ (Burkardt-Cottingham)} sum rule \cite{BurCot70}.
\begin{equation}
 S_{{\rm BC}} =
\int_0^1 dx \; G_2(x) = 0
\end{equation}
\section{Sum Rules in QCD}
In the full
 theory of strong interactions (QCD), the Naive Parton Model and its
expectations for the values of the sum rules are 
modified. These
 modifications are of two types. At high energy (large momentum
transfers), the 
coupling strength  of QCD becomes small and perturbation theory
can be used \cite{politzer,gross-wil}. In this regime, corrections to the 
sum rules can be expressed as 
a power series expansion in the strong coupling constant $\alpha_s(Q^2)$.  
At lower values of $Q^2$ non-perturbative corrections enter which can be
expressed as power series in $1/Q^2$. Unlike the perturbative corrections,
these cannot 
be calculated at present. In some cases, corrections to different
processes can be related to each other and experimental results
may be used to
determine 
the effect of these corrections on the sum rules. This section
analyses both the perturbative and non-perturbative corrections.

In Section 2.1  perturbative corrections are given first
 in the framework of the QCD improved
Parton Model. This approach allows
to give a very appealing and intuitive
description of the
  basic ideas of factorization and $Q^2$ evolution
of structure functions. 
We restrict our discussion in this paragraph to leading order
corrections and will see
 that in the leading logarithmic approximation
 all sum rules remain valid.

For the discussion of higher order perturbative corrections 
it is convenient to employ the framework of the  
operator product expansion, since the 
structure of the corrections becomes most transparent   
in this more formal approach. It leads of course to the same
results as one would get with the QCD Parton Model.
In fact, the connections between both descriptions will be pointed
 out wherever possible. 

The operator product expansion has the further advantage that
non-\-per\-tur\-ba\-tive effects can easily be incorporated.
 Power corrections of higher twist
are studied in Section 2.2.
\subsection{\it Perturbative QCD Corrections}
\subsubsection*{\sc QCD Parton Model}
In the previous chapter the relation between 
the structure functions and quark distributions was given and it
was stated that the Naive Parton Model predicts 
the $Q^2$-in\-de\-pen\-dence of the structure functions $F(x,Q^2)$.
Violations of this scaling behaviour were observed experimentally
and may be explained theoretically due to strong interactions.
The QCD improved Parton Model gives a simple and quantitative 
description of these effects and introduces the correct
  $Q^2$
dependence into the parton distribution functions. 

The QCD generalization of Eqn. (\ref{f2parton}) is provided by the
factorization theorem for deep inelastic scattering
(see \cite{ColSopSte89,QCDhandbook})
\begin{equation}
\label{factorization}
{\cal F}_k^{{\rm NS}}(x,Q^2) = 
\int_0^1\frac{dy}{y}
\tilde{C}_k \left(
  \frac{x}{y},\frac{Q^2}{\mu^2},\frac{\mu_f^2}{\mu^2},\alpha_s(\mu^2)
            \right)
  f_k^{{\rm NS}}(y,\mu_f,\mu^2)
.\end{equation} 
Here ${\cal F}_k$ serves as a generic notation for $2F_1,F_2/x$ and $F_3$.
For polarized structure functions 
${\cal F}_k=2G_1,G_2$ one should replace $f_k$ by $\Delta f_k$.
For simplicity we consider the nonsinglet combination of structure
 functions like $ep-en$. Accordingly $f_k^{{\rm NS}}$ are the 
appropriate combinations of parton densities, e.g. for 
${\cal F}_2^{{\rm NS}}=F_2^{{\rm NS}}/x$ one has
$f_2^{{\rm NS}}=\sum_i q_i^2 f_i^{{\rm NS}}.$
In Eqn.(\ref{factorization})  the factorization of
 high momentum (short distance) and
low momentum (long distance) effects
is expressed. The former are described by
the coefficient function $\tilde{C}_k$ and calculable in perturbation
theory. As a characteristic feature for perturbative computations
one finds a dependence on the renormalization scale $\mu^2$.
Long distance contributions cannot be calculated by  present theoretical
methods available
 in QCD and are absorbed in the parton distribution functions.
The separation between the low and high momentum regime calls for
another scale,
 the factorization scale $\mu_f$. In the following
we shall choose $\mu_f^2=\mu^2$.

The coefficient functions of 
Eqn.(\ref{factorization}) were calculated in the leading
logarithmic approximation in 
\cite{AltPar77}, to
order $\alpha_s$ in \cite{AltEllMar78,HumNee81} and order
$\alpha_s^2$ in 
\cite{NeeZij91a,ZijNee91b,ZijNee92a,ZijNee92b,ZijNee94}.
In leading order QCD 
they have the following 
form (in this and all
 subsequent equations $\alpha_s$ without an argument is understood to mean 
$\alpha_s(Q^2)$)
\begin{equation}
\tilde{C}_k^{{\rm NS}}
=
\delta\left(1-\frac{x}{y}\right)
+
\frac{\alpha_s}{2\pi}\left[
  P_{qq}\left(\frac{x}{y}\right)\ln\frac{Q^2}{\mu^2}
 +R^{{\rm NS}}_k\left(\frac{x}{y}\right)
                          \right]
\end{equation}
The well known splitting function 
$P_{qq}(z)=C_F[(1+z^2)/(1-z)_+ + (3/2)\delta(1-z)]$
measures the variation with $Q^2$ of the probability
of finding a quark inside a quark with a fraction
$z=x/y$ of its momentum y \cite{AltPar77}.
This leading logarithmic term
is universal for all structure functions and can be absorbed 
into newly defined, $Q^2$-dependent parton distribution functions
\begin{equation}
\label{Q2PDF}
f_k^{{\rm NS}}(x,Q^2) = f_k^{{\rm NS}}(x)+
\frac{\alpha_s}{2\pi}
\int_0^1\frac{d\xi}{\xi}f_k^{{\rm NS}}(\xi)
\left[
  P_{qq}\ln\frac{Q^2}{\mu^2}
 +R^{{\rm NS}}_{k,{\rm abs}}
                          \right]
.\end{equation}
The particular way  how 
$R^{{\rm NS}}_{k}$ is split into an absorbed part
$R^{{\rm NS}}_{k,{\rm abs}}$ and a remaining part
$R^{{\rm NS}}_{k,{\rm rem}}$ is a matter of convention and
 specifies the so-called factorization scheme.
Two popular choices are the DIS scheme and the $\msbar$
factorization scheme (for details see e.g. \cite{QCDhandbook}).
With the $Q^2$ dependent quark distributions 
the structure functions can be rewritten in the following form 
if terms of order $\alpha_s^2$ are neglected:
\begin{equation}
\label{factorLLA}
\ba{ll}\dsp
{\cal F}_k^{{\rm NS}}(x,Q^2)
& \dsp
 = 
\int_0^1\frac{dy}{y}
\Bigg(
   \delta\left(1-\frac{x}{y}\right)
   +\frac{\alpha_s}{2\pi}
     R^{{\rm NS}}_{k,{\rm rem}}
\Bigg)
f_k^{{\rm NS}}(y,Q^2)
\\ & \dsp
\stackrel{LLA}{\longrightarrow}
f_k^{{\rm NS}}(x,Q^2) + {\cal O}(\alpha_s)
\ea
\end{equation}
It can be seen in the last step that 
in the leading logarithmic approximation (LLA) 
the relations between structure functions and 
parton distribution functions remain unchanged except 
that the parton densities now depend on $Q^2$.
With the same modification all Parton Model sum rules
  of the previous chapter
remain valid in this approximation.

The $Q^2$ dependence of 
the distribution functions and hence 
the structure functions and sum rules
 is most readily expressed
by the 
Dokshitzer-Gribov-Lipatov-Altarelli-Parisi (DGLAP) \cite{dglap,AltPar77}
evolution equation which to leading order in QCD has the following form
\begin{equation}
Q^2\frac{df_i(z,Q^2)}{dQ^2}=\frac{\alpha_s}{2\pi}\int_z^1
(f_i(y,Q^2)P_{qq}(\frac{z}{y})+g(y,Q^2)P_{qg}(\frac{z}{y})){dy\over y}
.\end{equation}
If one is not restricted to flavour non-singlet combinations
the other function $P_{qg}(y)$ comes into play due to
the probability of finding a quark inside a gluon.  
Both splitting functions
are  determined
 by perturbative QCD and can
be written as an expansion in $\alpha_s$.
This equation can be used to determine 
the perturbative QCD corrections to the various 
sum rules.
 It should be noted that the DGLAP equation contains more information 
about the 
behaviour of  the structure functions than does the set of sum rules.
However, the higher order QCD corrections 
to the sum rules are easier to compute than  
those to the DGLAP equation. Hence while
the DGLAP evolution equation is only known to order $\alpha_s^2$ 
 \cite{Flo81a,Flo81b,Flo81c,CurFurPet80,Fur80a,Fur80b,lopez81},
\cite{ZijNee94,vog1,vog2,vog3},  
the corrections that we discuss below to some 
of the sum rules are known to order $\alpha_s^3$.
\subsubsection*{\sc Operator Product Expansion Approach}
Modifications of the 
 sum rules due to higher order QCD corrections
were indicated in the previous section
for the
QCD improved Parton Model. In this section we explicitly discuss 
those corrections.
 Their structure becomes particularly
  transparent in the 
framework of the operator product 
 expansion \cite{Wil69}.
This approach will  also prove to be useful in the following section
 for the discussion of
non-\-per\-tur\-ba\-tive effects. 
The Mellin moments of the structure functions
 are expanded in a form  \cite{ChrHasMue72}
 where the short distance and
long distance contributions are factorized
 in a similar fashion as in
Eqn.(\ref{factorization})
\begin{equation}
\int_0^1 dx \;
 x^{n-1} {\cal F}_k (x,Q^2)
 = \sum_{i,\tau}
 C^{i,\tau}_{k,n}\left(\frac{Q^2}{\mu^2},\alpha_s(\mu^2)\right)
 A^{i,\tau}_n(\mu^2).
\label{factor}
\end{equation}
The expansion is expressed
in terms of reduced matrix elements $A^{i,\tau}_n$ of operators 
renormalized at scale $\mu$ which
describe the long distance effects and have
to be determined from experiments and 
coefficient functions
$C^{i,\tau}_{k,n}$ that describe
 the short distance effects and can
be calculated perturbatively. The 
operators are characterized by their quantum numbers
which can be broken into two types.
The label $\tau$ refers to the twist of the operator
 and $i$ refers to the 
flavor quantum numbers such as isospin. 
The ``twist'' of an operator is defined by its
dimension $d_{{\cal O}}$
minus its spin $n$. Since $C^{i,\tau}_{k,n}\left(Q^2/\mu^2,
\as\right)\sim (\mu^2/Q^2)^{(\tau -2)/2}$, operators
of lowest twist ($\tau =2$) dominate in the large $Q^2$ limit.
 We shall postpone the discussion of higher twist operators
to the next section and
 omit the twist label when considering
the leading, twist-2, terms.

If one considers for simplicity the moments of
flavour non-\-singlet structure functions,  one can 
readily see
the similarity between the approaches of 
 the QCD Parton Model and the 
operator product expansion.
Taking the moments on both sides
 of Eqn.(\ref{factorization}) 
shows that the Parton Model 
analogues of the  coefficient functions and operator 
matrix elements of Eqn.(\ref{factor}) are 
given by the following moments
\begin{equation}
\ba{lll}\dsp
C_{k,n}^{{\rm NS}}
\left(\frac{Q^2}{\mu^2},\alpha_s(\mu^2)
\right)
& \Longleftrightarrow
& \dsp
\int_0^1 dx \;x^{n-1}\; \tilde{C}_k^{{\rm NS}}
\left(x,\frac{Q^2}{\mu^2},\alpha_s(\mu^2)
\right)
\\ \dsp
A_{n}^{{\rm NS}}(\mu^2)
& \Longleftrightarrow
& \dsp
\int_0^1 dx \;x^{n-1}\; f_k^{{\rm NS}}(x,\mu^2)
\ea
\end{equation}

The $Q^2$ behaviour of the coefficient functions is governed by
their renormalization group equation \cite{rgeq}, which follows
from the fact that the LHS of Eqn.(\ref{factor}) as a 
measurable quantity is independent of $\mu^2$.
The non-\-singlet operators are renormalized
 multiplicatively with renormalization constant $Z_n^{\rm NS}$.
The renormalization group equation
then has the
following simple form 
\begin{equation}\label{RGE}
\Bigg[
\mu^2\frac{\partial}{\partial \mu^2}
 + \beta \frac{\partial}{\partial A}
+\gamma_n^{{\rm NS}}
\Bigg]
C_{k,n}^{{\rm NS}} \left(\frac{Q^2}{\mu^2},
    A(\mu^2)\right)
= 0
,\end{equation}
where
$A\equiv \alpha_s/\pi$.
 The 
 beta-function and the anomalous dimension are defined by
\begin{equation}
 \beta\big(A(\mu^2)\big)=\mu^2\frac{d}{d\mu^2}A(\mu^2),\;\;\;
\gamma_n^{{\rm NS}}=\mu^2\frac{d}{d\mu^2}
 \log Z_n^{{\rm NS}}
\end{equation}
and are given by the expansions
\begin{equation}
\ba{rl}\dsp
\beta(A) =
& \dsp
-\beta_0 A^2-\beta_1 A^3
-\beta_2 A^4 +\dots
\\ \dsp
\gamma_n^{{\rm NS}}(A)=
& \dsp
\gamma_n^{{\rm NS}(0)} A
+\gamma_n^{{\rm NS}(1)} A^2
+\gamma_n^{{\rm NS}(2)} A^3
+\dots
.\ea
\label{anom-gam}
\end{equation}
The solution of Eqn.(\ref{RGE}) is
\begin{equation}\label{RGEsolution}
C_{k,n}^{{\rm NS}}\left(\frac{Q^2}{\mu^2},
A(\mu^2)\right)
=
C_{k,n}^{{\rm NS}}\left(1,\bar{A}(Q^2)\right)
  \exp\left( \int_{\bar{A}(\mu^2)}
                 ^{\bar{A}(Q^2)}dx
              \frac{\gamma_n^{{\rm NS}}(x)}{\beta(x)}
      \right)
\end{equation}
where the formula for the effective coupling constant
 $\bar{\alpha_s}$ and the coefficients
of the anomalous dimension 
and the beta-function are listed
 in the appendix.

From Eqn.(\ref{RGEsolution}) it becomes apparent 
that  QCD 
corrections in $C_{k,n}^{{\rm NS}}$
 are twofold. They
arise from the expansion of
\begin{equation}\label{CFexpansion} 
C_{k,n}^{{\rm NS}}\left(1,\bar{A}(Q^2)\right)
=
1+B_{k,n}^{{\rm NS}(1)}\bar{A}(Q^2)
+B_{k,n}^{{\rm NS}(2)}\bar{A}^2(Q^2)+\dots
\end{equation}
as well as from the expansion of the exponential
on the RHS of Eqn.(\ref{RGEsolution}). The $Q^2$
dependent part of that  term is 
\begin{equation}\label{Q2exponential}
\ba{l}\dsp
\exp\left(
\int^{\bar{A}(Q^2)}dx\frac{\gamma_n^{{\rm NS}}(x)}{\beta(x)}
\right)
=
  \left[\bar{A}(Q^2)
  \right]^{-\frac{\gamma_n^{{\rm NS}(0)}}{\beta_0}}
\\ \dsp
\hphantom{xxxxxxx}\cdot
\Bigg\{
1+\bar{A}(Q^2)\left[
-\frac{\gamma_n^{{\rm NS}(1)}}{\beta_0}
+\frac{\beta_1\gamma_n^{{\rm NS}(0)}}{\beta_0^2}
             \right]
+ \bar{A}^2(Q^2)[\dots ]
\Bigg\}
\ea
\end{equation}
The leading  term 
$\gamma_n^{{\rm NS}(0)}$
of the anomalous dimension is
independent of the renormalization scheme.
This is no longer the case for the higher order terms
 $\gamma_n^{{\rm NS}(i)},i>0$. 
However, in  expressions for physical quantities
they are  associated with the scheme dependent
coefficients $B_{k,n}^{(i)}$ in such a way that 
the final answer is renormalization scheme invariant.

The $\mu^2$ dependent part of the exponential
in Eqn.(\ref{RGEsolution}) may be combined
 with the 
reduced matrix element $A_n^{{\rm NS}}(\mu^2)$
to form the renormalization scale
 invariant expression
\begin{equation}\label{Ainv}
A_n^{{\rm NS, inv}}=
\exp\left(
-\int^{\bar{A}(\mu^2)}dx\frac{\gamma_n^{{\rm NS}}(x)}{\beta(x)}
\right)
A_n^{{\rm NS}}(\mu^2)
.\end{equation}
In the case of the sum rules all relevant operators 
have 
$\gamma_{n=1}^{(0)}=0$. This can be observed from the 
relation $\int_0^1 dx \; x^{n-1}\; P_{qq}(x)=
\gamma^{{\rm NS}(0)}_n/2$, which vanishes for $n=1$ as
as consequence of fermion number conservation.

 In view of
 Eqns. (\ref{CFexpansion},
\ref{Q2exponential})
the RHS of Eqn.(\ref{factor}) approaches 
a constant value  as $Q^2\rightarrow \infty$ that is basically
given by
 $A_n^{{\rm NS, inv}}$
 and may 
  be identified with the corresponding expression obtained
in the Naive Parton Model.
As will be seen below the situation becomes even simpler, when the 
operators under consideration are conserved currents.
In this case the anomalous dimensions vanish and the
  QCD corrections are already 
completely determined through Eqn.(\ref{CFexpansion}). 
Since these operators are not renormalized, their reduced
matrix elements $A_n^{{\rm NS}}$ are independent of $\mu^2$.

Analogous relations to Eqns. (\ref{RGE},\ref{RGEsolution},\ref{Ainv})
 hold for singlet combinations of structure 
functions.
They are more complex than for the nonsinglet
case, 
since mixing of different operators with the  same 
quantum numbers may occur under renormalization,  leading
to an anomalous dimension matrix. 

Let us illustrate the above discussion in an example
and consider the moments of the structure functions
$F_2$ and $G_1$ of deep inelastic electron-nucleon
scattering
\begin{equation}\label{moments}
\ba{lll}\dsp
\int_0^1 dx \;x^{n-2}\; F_2^{ep-en} 
& \dsp 
= \sum_i C_{F_2,n}^i v_i^{{\rm NS}}

& \dsp
\;\;\;n=2,4,6\dots
\\ \dsp  
\int_0^1dx \; x^{n-1}\; G_1^{eN} 
& \dsp 
= \frac{1}{2}\sum_i C_{G_1,n}^i a_i^N
& \dsp
\;\;\;n=1,3,5\dots
\ea
\end{equation}
which leads to the 
{\sc Valence Isospin sum rule} and the 
{\sc Spin sum rules} respectively.
Depending on the 
crossing properties of the  structure functions
under $\mu \leftrightarrow \nu, x\leftrightarrow -x$,
only operators with  definite spin signatures 
are relevant
in the expansions of Eqn.(\ref{moments}).
Spin-even operators contribute to the moments 
of the combination
$F_2^{ep-en}$ and  
spin-odd operators  to those of
$G_1^{eN}$. For later use we note that 
$F_{1,2}^{\nu n-\nu p},F_3^{\nu n +\nu p}$
also represent spin-odd combinations.

It is obvious that the {\sc Spin sum rules} are 
immediately obtained from the first moment $n=1$ of
$G_1$ in Eqn.(\ref{moments}).
In this case the corresponding operators
are the flavour
nonsinglet and singlet  
axial vector currents
\begin{equation}
J_{5\mu}^j
= {\bf \bar{\psi}} \gamma_{\mu} \gamma_5
                              \left(
  \frac{{\bf \lambda}_j}{2}    \right)
  {\bf \psi}
\;\;\; ; \;\;\;
J_{5\mu}^0
= {\bf \bar{\psi}} \gamma_{\mu} \gamma_5
  {\bf \psi}
\end{equation}
 of the $SU(3)_F$
symmetry group $(j=1,\dots,8)$ 
(the $\lambda_j$ are the Gell-Mann matrices).  
The $a_i^N$ are given by the matrix
elements of these operators between 
the states of the nucleon $N=p,n$ with
momentum $P_{\mu}$, spin $S_{\mu}$ and mass $M$ 
\begin{equation}\label{matrixelement}
\langle P,S | J_{5\mu}^j | P,S \rangle
= M a_j^N S_{\mu}
;\;\; ; \;\;\;
\langle P,S | J_{5\mu}^0 | P,S \rangle
= 2M a_0^N S_{\mu}
\end{equation}
The nonsinglet 
  axial vector currents are conserved in the massless quark 
limit. According to the discussion above the 
$a_j^N$ are  independent of $\mu^2$  and the 
corresponding anomalous dimension vanishes
$\Delta \gamma_{n=1}^{{\rm NS}}=0$. (We use the notation
$\Delta \gamma_n$ for  polarized scattering in distinction
to $\gamma_n$ for unpolarized scattering.)
The scale dependence of $a_0(\mu^2)$ on the other hand 
reflects the fact that the singlet axial vector current
is not conserved due to the axial anomaly
\cite{Adl69,BelJac69} and therefore
has a nonvanishing anomalous dimension 
$\Delta \gamma_{n=1}^{{\rm S}}\neq 0$. 
The polarized anomalous dimension
$\Delta\gamma_n^{{\rm NS/S}}$ 
for arbitrary $n$ 
was calculated in
leading order
in \cite{AhmRos75,Sas75,AltPar77}
and next-\-to-\-leading order in
\cite{Kod80,ZijNee94,MerNee95}. 
The next-\-next-\-to-\-leading order  result
$\Delta \gamma_{n,qq}^{{S}(2)}$
can be found in
\cite{Lar93,CheKue93}.
The nonleading results were calculated 
using the $\msbar$ scheme, 
 the standard modification
of the Minimal Subtraction scheme \cite{tHo73}.
(A noteworthy feature of this particular renormalization scheme
is the separate  gauge invariance of both the anomalous dimensions
 and the 
coefficient functions \cite{CasWil74}.) 
Some of them 
are listed in the appendix.
The Naive Parton Model expressions for the
 quantities $a_0^{{\rm inv}}$, $a_3$ and
$a_8$ as well as their values were
already  presented
 in Eqn. (\ref{aeqn}).

The situation for unpolarized 
lepton-nucleon scattering is not as straightforward
as for the polarized case. Since only spin-even
operators contribute to the operator product expansion
for the moments of $F_2$ in Eqn.(\ref{moments}), the 
{\sc Valence Isospin sum rule}
 cannot be obtained simply
as the special case $n=1$ of the first moment.
Nevertheless, the information that is contained in 
Eqn.(\ref{moments}) can  be used to derive QCD corrections to 
the {\sc Valence Isospin sum rule}.
The quantities
$v_j^{{\rm NS}}$ are   
defined as the reduced matrix elements of operators of the form 
${\cal O}_j^{{\rm NS}\mu_1,..,\mu_n}=
\left\{
\bar{\psi}\gamma^{\mu_1}D^{\mu_2}..D^{\mu_n}
(\lambda^j/2)\psi
\right\}_{S\{\mu\}}$ with $n=2,4,6,\dots$ and where $S$
indicates symmetrization of the indices.
The perturbative expansions
 for the corresponding anomalous
dimension $\gamma_n^{{\rm NS}}$ and the 
coefficient function $C_{F_2,n}^{{\rm NS}}(1,\bar{A}(Q^2)$
 are a priori meaningful only for even $n$. 
However, the QCD Parton Model not only
reproduces the results of the operator product expansion,
but also provides an answer for  moments with odd $n$, for
which no operators are available. QCD corrections to the 
{\sc Valence Isospin sum rule} may therefore be obtained by 
analytically continuing the results valid for even $n$ to 
the formally forbidden  values of odd $n$.
Similarly a continuation for spin odd combinations
of structure functions to values of even $n$ can be made.
Consequently the generalization of the anomalous dimensions
$\gamma_{n={\rm even/odd}}$ are denoted as
$\gamma_n^{\pm}$, now valid for all $n$.     
In leading order the nonsinglet and singlet
anomalous dimensions 
$\gamma_n^{{\rm NS/S}(0)}$
were obtained in
\cite{GeoPol74,GroWil74,AltPar77}.
The nonsinglet (singlet) 
 next-\-to-\-leading order 
$\msbar$ result
$\gamma_n^{{\rm NS}(1)\pm}$  
($\gamma_{n,ij}^{{\rm S}(1)}$)  
 was calculated 
in 
\cite{FloRosSac77}
(\cite{FloRosSac79,CurFurPet80,HamNee92})
 and simplified in
\cite{GonLopYnd79} 
(\cite{GonLop80}).
Furthermore, 
the anomalous dimensions are given by the moments
of
the splitting functions $P_{ij}^{(1)}$, which
were directly computed in
 \cite{Flo81a,Flo81b,Flo81c,CurFurPet80,Fur80a,vog1}.
Three loop nonsinglet anomalous dimensions
 $\gamma_n^{{\rm NS}(2)}$ are given in 
\cite{LarTkaVer91a,LarRitVer94} for even $n\geq 2$.
We note in passing that spin-even and spin-odd combinations
of structure functions correspond to combinations 
$q_f\pm \bar{q}_f$ and $\Delta q_f\mp \Delta \bar{q}_f$
of
parton densities
(see e.g. \cite{Alt82,GluReyStr95}).

We have demonstrated that QCD corrections to the
{\sc Valence Isospin sum rule}
can be derived
from the operator product expansion for the  moments
   of $F_2$, even though no operators exist for 
the first moment. That there is 
no corresponding operator for $n=1$ is reflected 
in the fact that  no 
reliable numerical value for $S_{{\rm G}}$ 
 can be given.  
Indeed, the Parton Model prediction for the  
{\sc Valence Isospin sum rule} was based on an 
additional assumption about the quark sea which has no
solid theoretical justification.

Within the same flavour
octet the  $Q^2$-evolution for all nonsinglet operators 
in Eqn.(\ref{factor}) 
is given by a common coefficient function defined by
$C_{k,n}^i=\delta_k^i C_{k,n}^{{\rm NS}}$, $i=1..8$.
Similarly, for the first moment one may 
 denote the singlet coefficient function
by
 $C_{k,n=1}^0 = \delta_k^0C^{{\rm S}}_k$.
The constants $\delta_k^i$ 
are combinations of quark charges
or weak couplings. In our example of the neutral current
($ J^{em}_{\mu}=\sum_f Q_f \bar{\psi}_f\gamma_{\mu}\psi$)
process one has
 $\delta_k^3=1/6,\delta_k^8=1/(6\sqrt{3}),\delta_k^0=2/9$ 
and all other $\delta_k^i=0$ for $k=F_2^{eN},G_1^{eN}$.
Next-to-leading order coefficient functions 
for the structure functions 
$F_1,F_2,F_3$ of
electron and neutrino scattering off nucleons
are given
for the $\msbar$ scheme
in 
\cite{BarBurDukMut78,FloRosSac79,AltEllMar79,HarKanSak79} 
 and for other schemes
in \cite{ZeeWilTre74,HinLle77,RujGeoPol77},
\cite{Cal77,AbaHum78,AltEllMar78,ross78,Kub79}.
For the various sum rules only the numerical values
of the corrections will be given in this section. The 
corresponding analytic expressions for the coefficient
functions and the anomalous dimensions can be found in the 
appendix.

For the  {\sc  Valence Isospin sum rule} 
\begin{equation}
\ba{ll}\dsp
S_{{\rm G}} 
= \frac{1}{3}
 C^{(ep-en)}_{F_2,1}(1,\bar{A}(Q^2)) 
  \exp\left(\int
                 ^{\bar{A}(Q^2)}dx
              \frac{\gamma_{n=1}^{{\rm NS}+}(x)}{\beta(x)}
      \right)
\ea
.\end{equation}
it was found \cite{BarBurDukMut78} that
 $B^{{\rm NS}(1)}_{F_2,n=1}=0$ vanishes.
The QCD corrections are therefore
due to the nonvanishing  anomalous dimension
$\gamma_{n=1}^{{\rm NS}(1)+}$:
\begin{equation}
\ba{rl}
S_G =
& \dsp \frac{1}{3}\left[
    1+\frac{1}{3}\frac{13+8\zeta(3)-2\pi^2}{33-2n_f}\apib
                 \right]
\\ = & \dsp
  \frac{1}{3}\left[
    1+ \left( {0.036 \atop 0.038}
       \right) \left(\apib\right)
                 \right]
\ea
\end{equation}
Here the upper and lower coefficients refer to 
$n_f=3$ and 4 respectively. 

For polarized electron-nucleon
scattering the moments read
\begin{equation}
\ba{ll}\dsp
\int_0^1 &\dsp dx     G_1^{ep(en)} (x,Q^2)
\dsp
= C^{{\rm NS}}_{G_1,1}(1,\bar{A}(Q^2)) 
 \frac{1}{12}\left(\pm a_3+\frac{1}{\sqrt{3}}a_8\right)
\\ & \dsp
+ C^{{\rm S}}_{G_1,1}(1,\bar{A}(Q^2)) 
  \exp\left( \int_{\bar{A}(\mu^2)}
                 ^{\bar{A}(Q^2)}dx
              \frac{\Delta\gamma^{{\rm S}}_{n=1,qq}(x)}{\beta(x)}
      \right)
 \frac{1}{9}a_0(\mu^2)
\\ &  \dsp
= C^{{\rm NS}}_{G_1,1}(1,\bar{A}(Q^2)) 
 \frac{1}{12}\left(\pm a_3+\frac{1}{\sqrt{3}}a_8\right)
\\ & \dsp
+ C^{{\rm S}}_{G_1,1}(1,\bar{A}(Q^2)) 
  \exp\left( \int^{\bar{A}(Q^2)}dx
              \frac{\Delta\gamma^{{\rm S}}_{n=1,qq}(x)}{\beta(x)}
      \right)
 \frac{1}{9}a_0^{{\rm inv}}
\ea
\end{equation}
In the nonsinglet part of this 
equation the exponential does not 
contribute because of the vanishing
anomalous dimension 
$\Delta\gamma_{n=1}^{{\rm NS}}=0$.
In the singlet part the renormalization scale dependent 
piece of the exponential may be  combined with the 
operator matrix element to form the scale
independent quantity $a_0^{{\rm inv}}$.
The corrections to the nonsinglet coefficient function
$C^{{\rm NS}}_{G_1,n=1}$ 
were calculated to
orders $\as$ \cite{KodMatMutSasUem79},
$\as^2$ \cite{GorLar86,GorLar87,ZijNee94} 
and $\as^3$
\cite{LarVer91} and for the singlet coefficient
function 
$C^{{\rm S}}_{G_1,n=1}$ 
to order $\as$ \cite{Kod80} and $\as^2$
\cite{ZijNee94,Lar94}. 
The QCD corrections for the {\sc Polarized Isospin sum rule}
are  completely determined by the
nonsinglet coefficient function
$C^{{\rm NS}}_{G_1,n=1}$. 
 The numerical result reads
\begin{equation}
S_{{\rm Bj}}= \frac{1}{6}a_3\left[
 1-\apib
-\left({3.583 \atop 3.250}\right)
\left(\apib\right)^2
-\left({20.215\atop 13.850}\right)
\left(\apib\right)^3
                             \right]
\end{equation}

The {\sc Spin sum rules} are obtained 
by inserting the coefficients
and anomalous dimension of the appendix and read
\begin{equation}
\ba{l}\dsp
S_{{\rm EJ}}^{p(n)}
 =
 \frac{1}{12}
 \left[
 1-\apib
-\left({3.583 \atop 3.250}\right)
\left(\apib\right)^2
-\left({20.215\atop 13.850}\right)
\left(\apib\right)^3
              \right]\cdot
\\ \dsp
\hphantom{x}
\cdot
\left(\pm a_3+\frac{1}{\sqrt{3}}a_8\right)
+\frac{1}{9}a_0(Q^2) \left[
 1-\apib
-\left({1.096\atop -0.067}\right)
\left(\apib\right)^2
              \right]
\\ \dsp
\hphantom{x}
 =
 \frac{1}{12}\left[
 1-\apib
-\left({3.583 \atop 3.250}\right)
\left(\apib\right)^2
-\left({20.215\atop 13.850}\right)
\left(\apib\right)^3
              \right] \cdot
\\ \dsp
\hphantom{x}
 \left(\pm a_3+\frac{1}{\sqrt{3}}a_8\right)
+\frac{a_0^{{\rm inv}}}{9} \left[
 1
-\left({0.333 \atop 0.040 }\right)
\apib
-\left({0.550 \atop -1.082 }\right)
\left(\apib\right)^2
              \right]
\ea
\label{eljaffcval}
\end{equation}
where in the first step the choice $\mu^2=Q^2$ was made.

The validity of the  {\sc $G_2$ sum rule} 
can be derived from the nonexistence of leading twist 
operators for $n=1$ \cite{Jaf90,AnsEfrLea95}
in the 
operator product expansion of the  moments of $G_2$.
 A continuation from higher $n$ to $n=1$, similar to the
case of the {\sc Valence Isospin sum rule}, also leads
to a vanishing result due to a kinematical factor
$(n-1)/2n$ in the expansion
(see \cite{KodMatMutSasUem79}).
This result is confirmed by explicit calculations
in \cite{AltLamNasRid94,KodMatUem95}, despite an earlier contrary
claim \cite{MerNee93}:
\begin{equation}
S_{{\rm BC}} = 0
\end{equation}

We now turn to the situation for neutrino or antineutrino
scattering on protons. 
The combination $F_3^{\nu p}+F_3^{\nu n}$ 
in the {\sc Baryon sum rule}
transforms as a flavour
singlet.
 In view of our earlier discussion
one could  expect 
  a more complex renormalization
group equation 
than Eqn.(\ref{RGE}) for
$C^{(\nu {\cal N})}_{F_3,n=1}$. 
 However, in this case the gluon field
operator transforms differently under charge conjugation
than the quark field singlet operator and therefore no
mixing occurs.
Since the anomalous dimension
vanishes, the QCD corrections
are due to the coefficient function
$C^{(\nu {\cal N})}_{F_3,n}$
 \cite{BarBurDukMut78,LarVer91,ZijNee92b,ZijNee94}:
\begin{equation}
S_{{\rm GLS}}= 3\left[
 1-\apib
-\left({3.583\atop 3.250}\right)
\left(\apib\right)^2
-\left({18.976 \atop 12.198 }\right)
\left(\apib\right)^3
                 \right]
\end{equation}
It was pointed out in \cite{GorLar86,GorLar87,ZijNee94} that
the corrections up to second order $\alpha_s^2$
 for the {\sc Baryon sum rule} are identically the same 
as for the {\sc Polarized Isospin sum rule}. 
As may be seen from the explicit formulae in the appendix,
 corrections
for both sum rules differ at order $\as^3$ due to terms
$\sim d^{abc}d_{abc}$, representing diagrams with an internal
fermion loop and a purely gluonic intermediate state.
 
For the {\sc Isospin sum rule}
both the anomalous dimension
and the corrections to the 
 coefficient
function 
 $B^{(\nu n-\nu p)}_{F_2,n=1}$
up to order $\alpha_s^2$ were found
to vanish 
\cite{BarBurDukMut78,ZijNee92a}.
Therefore the {\sc Isospin sum rule} is  not violated:
\begin{equation}
S_A = 2
\end{equation}
Since this sum rule was derived by the use of
current algebra methods \cite{Adl66}, it is expected to
hold true in QCD for all orders in $\alpha_s$. 

The {\sc Second Isospin sum rule}
is also characterized by its
 vanishing anomalous dimension 
$\gamma^{{\rm NS}}_{n=1}=0$,
but has a nonzero
 coefficient function 
\cite{BarBurDukMut78,CheGorLarTka84,LarTkaVer91}.
As a result one has
\begin{equation}
S_{{\rm Bj}}^{{\rm unpol}}=
 1-0.667\apib
-\left({2.944 \atop 2.648} \right)
\left(\apib\right)^2
-\left({18.596 \atop 13.381}\right)
\left(\apib\right)^3
\end{equation}

Combining the {\sc Isospin sum rule} and the
{\sc Second Isospin sum rule}
leads to the corrections for the first 
 moment of the {\sc  Callan-Gross Relation}
for neutrino scattering.
\begin{equation}
\int_0^1 \frac{dx}{x} F_L^{\nu n-\nu p}
=\int_0^1 \frac{dx}{x} 
(F_2^{\nu n-\nu p}-2xF_1^{\nu n-\nu p})
=S_{{\rm A}}-2S_{{\rm Bj}}^{{\rm unpol.}}
\end{equation}
The {\sc Callan Gross Relation} for electron-nucleon
scattering is violated in QCD as well. 
The coefficient functions 
 of the longitudinal structure function
were calculated to leading order in
\cite{ZeeWilTre74,Cal77,NanRos75,HinLle77,RujGeoPol77}. 
The results for next-\-to-\-leading order corrections 
 turned out to be very controversial. 
Nonsinglet or singlet coefficient functions 
$\tilde{C}_L(x,Q^2)$ and their  moments
$C_{L,n}(Q^2)$ were
given in order $\alpha_s^2$ in
\cite{Her86,Mir87,Kaz90,NeeZij91a,ZijNee91b,ZijNee92a}
(see also \cite{San91})
and \cite{Duk82,Cou82,Dev84,Kaz88,LarVer93a}
respectively. There is complete agreement between 
\cite{NeeZij91a,ZijNee91b,ZijNee92a} and \cite{LarVer93a}.
Third order ${\cal O}(\alpha_s^3)$ corrections  
to the second and higher (even) moments of the nonsinglet
longitudinal structure function can be found in
\cite{LarTkaVer91a,LarRitVer94}. The corrections to 
the {\sc Callan Gross Relation} enter the cross-section
ratio $R=\sigma_L/\sigma_T$ which is measurable in
experiments.

  There have been attempts to estimate even higher order corrections.
  Reference \cite{Kat94a} estimates the $O(\alpha_s^4)$ 
  coefficients for the {\sc Second Isospin sum rule} and the 
  nonsinglet part of the {\sc Spin sum rules}. Neglecting light-by-light
  diagrams, this estimate is also used for the {\sc Baryon  sum rule}.
   These results are in reasonable agreement
  with another estimate by \cite{SamLiSte94}
  using Pad\'{e} approximation.
  Reference \cite{Kat94b} gives the $O(\alpha_s^3)$ estimate
  for the singlet part of the {\sc Spin sum rules}.
\subsection{\it Higher Twist Effects}
So far only leading twist corrections to the sum rules
have been discussed  and nucleon mass effects were neglected.
 Contributions of higher twist operators
in the operator product expansion of the structure function
 moments are suppressed by powers of $1/Q^2$.
These so-called dynamical power 
corrections are of non-\-perturbative nature
and may become important at low $Q^2$.
They are difficult to estimate, because at present no 
method is available for reliably calculating in the non-perturbative
regime.
Estimates of higher twist effects are therefore usually
accompanied by large uncertainties.

Another class of power corrections are 
of purely kinematical origin and arise in case that 
the nucleon mass is not neglected. Such {\it target mass} effects 
are suppressed by powers of $M^2/Q^2$ and may also become
relevant for low values of $Q^2$.
 They can be taken into
account to all orders in  
 $(M^2/Q^2)^n$ by replacing the 
structure function moments through so-called
Nachtmann moments 
\cite{Nac73}.

 A standard approach 
 \cite{ShuVai82a,ShuVai82b}
to calculate higher twist contributions
is based on 
 QCD sum rules
\cite{ShiVaiZak79}.
This method parameterizes nonperturbative effects 
in terms
of quark and gluon condensates. These condensates are
matched with a 
hadronic representation obtained phenomenologically
through a dispersion integral with a spectral density 
given by resonance and continuum contributions that can
be extracted from data.
The matrix elements of the higher twist
 operators are therefore
expressed through condensates, the values of which,
however, are only known to 
10-20\% \cite{ShiVaiZak79}.

Power corrections $\sim 1/Q^2$ to the 
first moments of the polarized structure function $G_1$
are given by \cite{BalBraKol90} 
(see also \cite{AnsEfrLea95})
\begin{equation}\label{HTEJ}
\delta S_{{\rm EJ}}^{p,n}
= -\frac{1}{6Q^2} 
\left\{
\frac{5}{3}\langle\langle{\cal O}^{{\rm S}}_{p,n}\rangle\rangle
+\langle\langle{\cal O}^{{\rm NS}}_{p,n}\rangle\rangle
\right\}
+\frac{2}{9}\frac{M^2}{Q^2}\int_0^1dx\;x^2\; G_1^{p,n}
\end{equation}
where 
\begin{equation}
\langle\langle{\cal O}\rangle\rangle
= \frac{8}{9}\left[
\langle\langle U \rangle\rangle
-\frac{M^2}{4}
\langle\langle V \rangle\rangle
\right]
\end{equation}
and 
$\langle\langle U \rangle\rangle
,\langle\langle V \rangle\rangle
$
are  defined by
\begin{equation}
\ba{rl}\dsp
\langle P,S|U_\mu|P,S\rangle = 
&\dsp
2 M S_{\mu}\langle\langle U \rangle\rangle
\\ \dsp 
\langle P,S|V_{\mu\nu,\sigma}|P,S\rangle = 
&\dsp
2 M \langle\langle V \rangle\rangle
\left\{(S_{\mu}P_{\nu}-S_{\nu}P_{\mu})P_{\sigma})
\right\}_{S\{\nu,\sigma\}}
\ea
\end{equation}
as the reduced elements of the operators
\begin{equation}
\ba{rl}\dsp
U_{\mu}^{{\rm S/NS}} = 
& \dsp
g_s\left[
\bar{u}\tilde{G}_{\mu\nu}\gamma^{\nu}u
\pm \bar{d}\tilde{G}_{\mu\nu}\gamma^{\nu}d
\right]
\\ \dsp
V_{\mu\nu,\sigma}^{{\rm S/NS}} = 
& \dsp
g_s\left\{
\bar{u}\tilde{G}_{\mu\nu}\gamma_{\sigma}u
\pm \bar{d}\tilde{G}_{\mu\nu}\gamma_{\sigma}d
\right\}_{S\{\nu,\sigma\}}
\ea
\end{equation}
for the proton 
 ($u\leftrightarrow d$  leads to  the
corresponding neutron operators). Here
the strange quark contribution is neglected and
\begin{equation}
\tilde{G}_{\mu\nu} = \frac{1}{2}\epsilon_{\mu\nu\alpha\beta}
G_a^{\alpha\beta}\left(\frac{\lambda_a}{2}\right)
.\end{equation}
After some errors of \cite{ShuVai82b} were 
discovered in \cite{JiUnr94}, the analysis of higher twist 
effects in \cite{BalBraKol90} was updated (see  \cite{AnsEfrLea95}).
\begin{equation}
\label{HTresQCD}
\ba{ll}
\delta S_{{\rm EJ}}^p
 = 
& \dsp
-\frac{(0.02\pm 0.013){\rm GeV}^2}{Q^2}
\\ \dsp
\delta S_{{\rm EJ}}^n
 = 
& \dsp
-\frac{(0.005\pm 0.003){\rm GeV}^2}{Q^2}
\ea
\end{equation}
These values agree approximately with another analysis
employing the QCD sum rule approach 
\cite{SteGorManSchGre95,SteGorManSch95}.
They
 are, however, 
 considerably smaller than
those obtained in
\cite{AnsIofLea89,BurIof92,BurIof93}
which rely on the relation
between the first moment of the polarized 
structure function $G_1^{p,n}$ and the
 Gera\-simov-\-Drell-\-Hearn sum rule 
\cite{Ger66,DreHea66}
for forward scattering of real photon off nucleons.
A re-examination  of the QCD
sum rule method in \cite{RosRob94} arrives at the 
result
$\delta S_{{\rm EJ}}^p=-(0.029\;{\rm GeV}^2)/Q^2$ and 
$\delta S_{{\rm EJ}}^n=-(0.002\;{\rm GeV}^2)/Q^2$
 with uncertainties of
about 20\%.
A recent paper \cite{knew96} calculated the one-loop anomalous dimension
for the $\tau=4$ operators that allow the logarithmic corrections to
the $1/Q^2$ term to be computed.

Power corrections for the 
{\sc Polarized Isospin sum rule}
follow immediately from Eqn.(\ref{HTEJ}):
\begin{equation}\label{HTBj}
\delta S_{{\rm Bj}}
= -\frac{1}{6Q^2} 
\langle\langle{\cal O}^{{\rm NS}}_{p-n}\rangle\rangle
+\frac{2}{9}\frac{M^2}{Q^2}\int_0^1dx\;x^2\; G_1^{p-n}
\end{equation}
 The first term was estimated in \cite{BalBraKol90}
to be 
$-(0.09\pm 0.06) {\rm GeV}^2/(6Q^2)$.
The leading target mass corrections are given by the  
second term. As is discussed in
\cite{Wan77,MatUem80,KawUem95a,KawUem95b},
these
kinematical power corrections $\propto (M^2/Q^2)^n$
 may be summed
up to all orders by the use of Nachtmann moments 
\cite{Nac73}. 
The third moment of $G_1$ in the mass term of
 Eqn.(\ref{HTBj}) can be determined from data. 
The integral
amounts to
0.0168 at $Q^2=2 \; {\rm GeV}^2$ and
0.0130 at $Q^2=4.6\; {\rm GeV}^2$
\cite{EllKar93}. 
Target mass effects are therefore 
negligible for experimentally
accessible $Q^2$. 
One thus  arrives at  
\cite{AnsEfrLea95}
\begin{equation}
\delta S_{{\rm Bj}} \approx
-\frac{(0.015\pm 0.013){\rm GeV}^2}{Q^2}
\end{equation}
This result is comparable with a diquark model
estimation of
\cite{AnsCarLev95}.

Higher twist effects were also studied for neu\-tri\-no-\-nucleon 
scattering. In \cite{BraKol87}
both the 
{\sc Baryon sum rule}
(see also \cite{ChyKat92})
 and the {\sc Second Isospin
sum rule} are discussed. 
With the reduced matrix element 
\begin{equation}
\langle P|{\cal O}^{{\rm S/NS}}_{5\mu}|P\rangle
_{{\rm \scriptstyle \hspace{-4ex}spin}
  \atop{\rm \scriptstyle averaged}}
\equiv 2P_{\mu}\langle\langle {\cal O}^{{\rm S/NS}}_5
\rangle\rangle
\end{equation}
of the twist four operators operators
\begin{equation}
{\cal O}^{{\rm S/NS}}_{5\mu} = 
\bar{u}\tilde{G}_{\mu\nu}\gamma^{\nu}\gamma_5 u
\pm \bar{d}\tilde{G}_{\mu\nu}\gamma^{\nu}\gamma_5 d
\end{equation}
one has 
\begin{equation}
\ba{ll}\dsp
\delta S_{{\rm GLS}}
& \dsp
 = 
3\left(-\frac{8}{27}\right)
\frac{\langle\langle {\cal O}^{{\rm S}}_5\rangle\rangle}{Q^2}
\\ \dsp
\delta S_{{\rm Bj}}^{{\rm unpol.}}
& \dsp
 = 
\left(-\frac{8}{9}\right)
\frac{\langle\langle {\cal O}^{{\rm NS}}_5\rangle\rangle}{Q^2}
\ea
.\end{equation}
Target mass corrections may again be summed to all orders
through the use of Nachtmann moments as discussed in
\cite{BraKol87} and
are very small.
The values for the reduced matrix elements were estimated
as \cite{BraKol87}
\begin{equation}
\ba{ll}\dsp
\langle\langle {\cal O}^{{\rm S}}_5 \rangle\rangle 
& \dsp
= 0.33\;
{\rm GeV}^2
\\ \dsp
\langle\langle {\cal O}^{{\rm NS}}_5 \rangle\rangle 
& \dsp
= 0.15\;
{\rm GeV}^2
\ea
\end{equation}
Within the large uncertainty of $\sim$ 50\% this value is 
in agreement with the  results from two other methods
also studied
in \cite{BraKol87}, namely the vector dominance approximation
and the nonrelativistic quark model. 
This indicates problems  in  bag-model calculations
\cite{FajOak85} which give negligibly small values
for twist four contributions. 
The reanalysis of \cite{RosRob94} presents 
as a result
$\langle\langle {\cal O}^{{\rm S}}_5 \rangle\rangle 
= 0.53\; {\rm GeV}^2$
with an error of about 20\%.

We are now in a 
position to compute the numerical values for the sum rules
and their various corrections.
\begin{table}
\begin{tabular}{|c||c|c|c||c|c|c|c|c|}
\hline
& $\dsp\frac{Q^2}{{\rm GeV}^2}$   &  $n_f$  
& $\bar{\as}^{(n_f)}(Q^2)$    
&  ${\cal O}(\as^0)$ 
&  ${\cal O}(\as^1)$ 
&  ${\cal O}(\as^2)$ 
&  ${\cal O}(\as^3)$ 
&  + HT
\\
\hline
$ S_{{\rm G}}$ 
& 4 & 4 & 0.315 
& 0.333 & 0.335 & - &-  &-  
\\
\hline
$ S_{{\rm GLS}}$ 
& 3 & 3 & 0.336 
& 3 & 2.571 & 2.583 & 2.486 & 2.388 
\\
\hline
$ S_{{\rm A}}$ 
& 3 & 3 & 0.336 
& 2 & 2 & 2 & 2 &  -
\\
\hline
$ S_{{\rm Bj}}^{{unpol.}}$ 
& 3 & 3 & 0.336 
& 1 & 0.905 & 0.902 & 0.872 & 0.828 
\\
& 10 & 4 & 0.258
& 1 & 0.930 & 0.931 & 0.920 & 0.907 
\\
\hline
$ S_{{\rm Bj}}$ 
& 2 & 3 & 0.385 
& 0.210 & 0.175 & 0.175 & 0.165 & 0.157
\\ 
& 3 & 3 & 0.336 
& 0.210 & 0.180 & 0.180 & 0.173 & 0.168 
\\
& 5 & 4 & 0.299 
& 0.210 & 0.184 & 0.185 & 0.181 & 0.178 
\\
& 10.4 & 4 & 0.256 
& 0.210 & 0.188 & 0.189 & 0.186 & 0.185 
\\
\hline
$ S_{{\rm EJ}}^p$ 
& 3 & 3 & 0.336 
& 0.185 & 0.165 & 0.166 & - & - 
\\
& 10.4 & 4 & 0.256
& 0.185 & 0.172 & 0.174 & - & -
\\
\hline
$ S_{{\rm EJ}}^n$ 
& 3 & 3 & 0.336 
& -0.024 & -0.0147 & -0.0146 & - & -
\\
& 10.4 & 4 & 0.256 
& -0.024 & -0.0154 & -0.0154 & - & - 
\\
\hline
$ S_{{\rm BC}}$ 
& 3 & 3 & 0.336 
& 0 & 0 & 0 & 0 &  -
\\
& 10 & 4 & 0.258
& 0 & 0 & 0 & 0 & -\cr\hline
\end{tabular} 
\caption[]{Theoretical estimates 
of the various sum rules. Corrections are taken 
into account up to and
including the order indicated in the columns.} 
\label{tablea}
\end{table}
As input parameter we take 
$\Lambda_{{\overline{MS}}}^{(5)}=233$ MeV, where
the superscript denotes the number of active flavors.
This value corresponds to $\alpha_s^{(5)}(M_Z^2)=0.12$.
The running coupling constant
 $\alpha_s^{(n_f)}(Q^2,\Lambda_{\overline{MS}}^{(n_f)})$
is calculated via the formula in the appendix.
The matching equation at $Q^2=M_q^2$
\begin{equation}
\begin{array}{ll}
\dsp  \alpha_s^{(n_f-1)}(M_q^2,\Lambda_{\overline{MS}}^{(n_f-1)})
=& \dsp
\alpha_s^{(n_f)}(M_q^2,\Lambda_{{\overline{MS}}}^{(n_f)})\cr
&\dsp \left[
1-\frac{7}{24}\left(
\frac{
\alpha_s^{(n_f)}(M_q^2,\Lambda_{{ \overline{MS}}}^{(n_f)})
}{\pi}
              \right)^2
\right]
\end{array}
\end{equation}
relates the coupling constant in the effective 
$n_f-1$ flavour theory to the full theory 
with
$n_f$ active flavors
\cite{BerWet82,Ber83,LarRitVer95}. It is implicitly solved 
in order to extract 
$\Lambda_{{ \overline{MS}}}^{(n_f)}$.
Applying this procedure subsequently at the thresholds
$Q^2=M_b^2=(4.7 \; {\rm GeV})^2$ and 
$Q^2=M_c^2=(1.6 \; {\rm GeV})^2$
 leads to 
$\Lambda_{{ \overline{MS}}}^{(4)}=320$ MeV and
$\Lambda_{{ \overline{MS}}}^{(3)}= 366$ MeV 
respectively.
The various
corrections  to the different sum rules are displayed in Table 1.
For the contributions of orders $\as,\as^2$ and $\as^3$
we used the 
the formula for the 
effective coupling constant including the  
leading, next-\-to-\-leading and 
next-\-next-\-to-\-leading terms respectively.\\
\section{Comparison of Theory  and Experiment}
\subsection{\it Experimental Issues and Results}
Experiments have been performed at the
 Stanford Linear Accelerator Center (SLAC) using polarized and unpolarized 
electron beams, at
FermiLab near Chicago using neutrino and unpolarized muon beams, 
at the European Center for Nuclear Research (CERN) in Geneva
using polarized and unpolarized muon beams and neutrino beams  and at
the 
Deutsches Elektronen-Synchrotron (DESY) in Hamburg using electron beams. 
Experiments are performed
 over a restricted range of $x$ and $Q^2$. Since the QCD corrections
to the sum rules depend on $Q^2$, data 
are required over the complete range of $x$ in as narrow a 
$Q^2$ range is practicable. The range of $x$
 is restricted to $x>Q^2/s$, In order for the QCD Parton Model
to make a reliable prediction $Q^2\gtap 2$ (GeV)$^2$,
 hence, for the sum rules to be measured,
data must be extrapolated into the very small $x$ region.
The extrapolation is least in experiments
 at the highest energy. To illustrate the extrapolation,
consider the {\sc baryon 
sum rule} measured in neutrino scattering. Figure 
2 shows  the CCFR data \cite{ccfr93}. 
\begin{figure}
\epsfbox{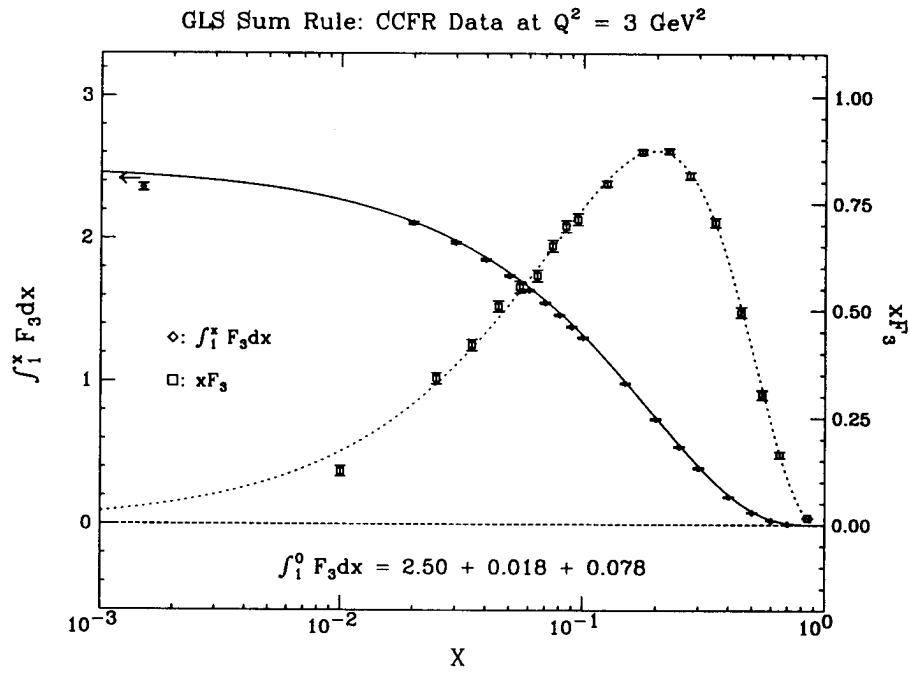}
\label{glls}
\caption[]{ The {\sc Baryon sum rule} from the CCFR 
experiment. Shown
 is $(xF_3^{ep}(x,Q^2)+xF_3^{en}(x,Q^2))/2$ (right hand scale)
and $\int_x^1dx(F_3^{ep}(x,Q^2)+F_3^{en}(x,Q^2))/2$ (left scale) 
at $Q^2=3 $ GeV$^2$. The curves are from a fit of the form $xF_3(x)
=Ax^b(1-x)^c$
}\end{figure}
As well as $F_3$, it shows $\int_{x_{min}}^1dx F_3(x)$
as a function of $x_{min}$; the 
lowest value of $x$ where data
 are available is  $x=0.015$. In performing the extrapolation to
$x_{min}=0$ a form 
for $xF_3(x)$ must be assumed. A fit of the form $xF_3(x)
=Ax^b(1-x)^c$ provides an excellent
 description of the data. A 
systematic error must be included in the quoted value of the sum
rule to take into account the
 extrapolation to $x_{min}=0$. 
This systematic error is difficult to estimate, since there is no
fundamental reason for preferring one extrapolation over another. 
Data from different values of $Q^2$ can only be combined if a $Q^2$
 extrapolation is assumed. Such an extrapolation can be based on
a fit to perturbative QCD. However, if this is done, a ``test of QCD''
 from the sum rules is compromised since 
perturbative QCD
 necessarily restricts the values that the sum rules can take.

Having given these caveats, we 
will now discuss the current experimental values for the sum rules. 
For the  {\sc baryon sum rule}, the CCFR collaboration \cite{ccfr93} gives, 
\begin{equation}
 S_{{\rm GLS}}=2.50\pm 0.018 (stat)\pm 0.078 (sys)
\end{equation} 
at $Q^2=3$ GeV$^2$. The systematic error
 includes the error from extrapolation into $x=0$.
A more precise result
 has been obtained by the CCFR collaboration \cite{harris95}
by  combining  their data with that from other experiments
 on neutrino scattering \cite{gargamelle,wa59,wa25,skat,fnal-e180}.
Also, in the very large $x$ region the nucleon's antiquark 
content is negligible and a relation between
$F_3^{\nu  p}$ and $F_2^{ep}$ exists in the Parton Model, {\it viz} \  
$5(F_3^{\nu p}(x,Q^2)+F_3^{\nu n}(x,Q^2))
=18(F_2^{e p}(x,Q^2)+F_2^{e n}(x,Q^2))$.  The experimental precision on the
{\sc baryon sum rule} accuracy can therefore  
be improved by including data from 
$F_2^{ep(n)}$ \cite{whitlow} which has much higher statistics. The combined 
data are then extrapolated
into the region below  $x=0.02$ and the sum rule evaluated.
Figure 
3 shows the extracted 
value of the {\sc baryon sum rule} as a function of $Q^2$. 
The curves on this figure will be discussed below.
\begin{figure}
\epsfbox{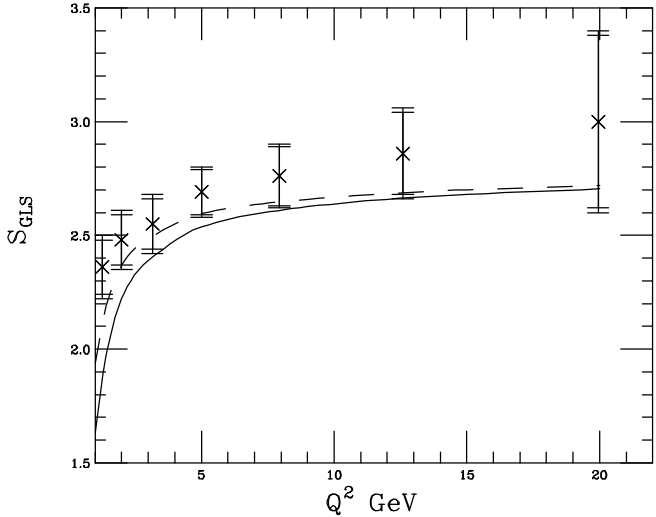}
\label{glls-1}
\caption[]{ The {\sc Baryon sum rule} from 
the CCFR experiment as a function of $Q^2$. The inner error bar
shows 
statistical errors and 
the outer a combination of statistical and systematic errors.
The curves are  a QCD predictions that are discussed in the text.}
\end{figure}
The solid line shows the QCD prediction including the higher twist
 effects, the dashed line shows the prediction of the higher
twist terms are ignored.  A comparison of the two curves shows 
that the higher twist contributions are unimportant for $Q^2\gtap 5$
GeV$^2$. The QCD prediction, which corresponds to $\alpha_s(M_Z)=0.12$, 
lies considerably below the data. The CCFR collaboration 
has fitted to QCD by allowing $\alpha_s$ to vary
 \cite{harris95} (see also \cite{ChyKat92,KatSid94,ChyRam95}) .
 The best fit corresponds to
\begin{equation}
\alpha_s(M_Z)=0.108 \pm 0.004 (stat) \pm 0.004 (syst)  \pm 0.006 (HT) 
\end{equation} 

The data for $F_2^{ep}-F_2^{en}$ and the integral 
$\int_x^1(F_2^{ep}-F_2^{en})dx/x$ are shown in Figure 
4
from \cite{nmc94}.
The  experiment observes no significant $Q^2$
 variation over the range $0.5 $ GeV$^2< Q^2<10$ GeV$^2$.
The  measured values are extrapolated into $x=0$ assuming 
 $F_2^{ep}-F_2^{en}=  ax^b$. $a$ and $b$  
are determined from a fit in the
region $0.004<x<0.015$ to be $a=0.2\pm 0.03$, $b=0.50\pm 0.06$. 
This extrapolation then contributes to the quoted error.
The value of the {\sc Valence Isospin sum rule} is
 determined by the NMC \cite{nmc94} collaboration to be 
\begin{equation}
S_G=0.235\pm 0.026
\end{equation}
for $<Q^2>=4$ GeV$^2$. This value is shown on the figure.
The same experiment has issued preliminary results from its 
full data set \cite{nmc95} which extends to  smaller values of $x$
and has fitted its data together with that of BCDMS
 \cite{bcdms} and SLAC \cite{whitlow} to give values of 
$F_2^{ep}$ and $F_2^{ed}$ ($d$ represents deuterium) over
 the x range $0.006<x<0.9$.  $Q^2$ corrections are 
applied to take 
higher twist effects into account and the results can then be interpreted
as \cite{eisele}
\begin{equation}
S_G=0.216\pm 0.027
\end{equation} 
for $<Q^2>=4$ GeV $^2$ with no significant remaining 
$Q^2$ dependence in the range $0.5<Q^2<10$ GeV$^2$.
Recent data from E665 \cite{e665} agree with NMC.
\begin{figure}
\epsfbox{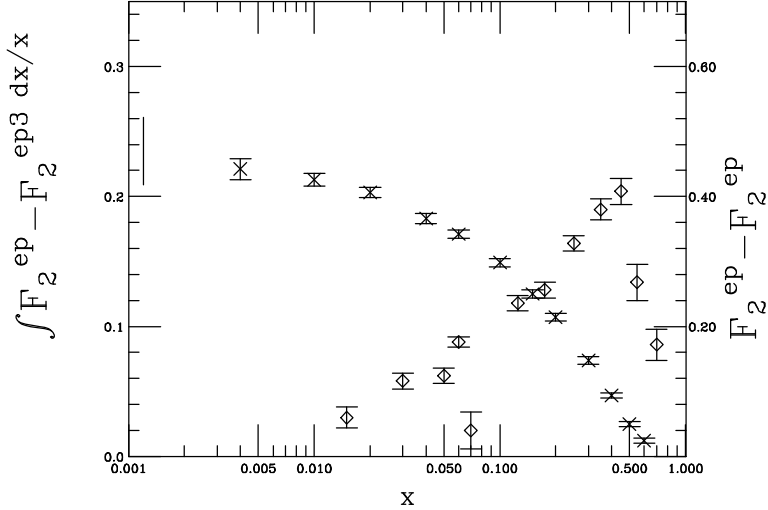}
\label{got-fig}
\caption[]{ Data from the NMC collaboration on
 the {\sc Valence Isospin  sum rule}. Shown is the quantity
 $F_2^{ep}-F_2^{en}$ (open points, right scale) and 
  $\int_x^1(F_2^{ep}-F_2^{en})dx/x$ (crosses, left scale).
The bar at the extreme left shows the derived value of the sum rule.}
\end{figure}

The {\sc Callan-Gross} relation is poorly
 determined. Experiments measure 
\begin{equation}
R(x,Q^2)=\frac{F_2(x,Q^2)(1+4M^2x^2/Q^2)-2xF_1(x,Q^2)}{2xF_1(x,Q^2)} 
\end{equation} 
Data from SLAC \cite{Bodek73,Miller72,Dasu88}  are shown in 
Figure 
5.
\begin{figure}
\epsfbox{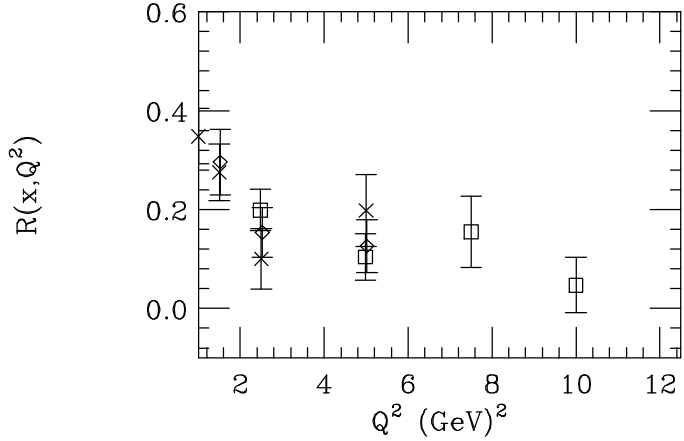}
\caption[]{ The quantity $R(x,Q^2)$ from the SLAC 
experiments as a function of $Q^2$ for certain
values of $x$ at $x=0.5$ 
(squares), $x=0.35$ (diamonds) and $x=0.2$ (crosses). }
\label{slac-r}
\end{figure}
The figure shows that $R(x,Q^2)$ falls
 rapidly as $Q^2$ is increased and that it is 
small. The value is consistent with that predicted by QCD. 
The {\sc Isospin} and {\sc Second Isospin sum rules}, 
which are related by the {\sc Callan-Gross} relation,
are difficult to measure with precision as they
require neutrino scattering off
hydrogen and deuterium targets and the statistical
 errors on such measurements are poor.
 Data show
no significant $Q^2$ variation in the range 
$2 \hbox{ GeV}^2 \ltap Q^2 \ltap 30$ GeV$^2$ and give \cite{alla}
\begin{equation}
S_{\rm A}=2.02\pm 0.40
\label{adler-value}
\end{equation} 
No published results are available on the 
{\sc Second Isospin sum rule}. However the
result of Eqn. (\ref{adler-value}) together
 with the information on $R(x,Q^2)$
show that the value of this rule is consistent with the expectation of
$S_{{\rm Bj}}^{{\rm unpol.}}=1$.

In the case of the {\sc Spin sum rules}, 
data are available from SMC at CERN \cite{SMC93,SMC94},
 using polarized muon beams scattering off  deuterium
and hydrogen targets, and E142 \cite{E14293}
and E143 \cite{E14395a,E14395b} at SLAC
 using polarized electron beams on He$^3$, 
 hydrogen  and deuterium targets targets.
The SLAC data only cover $x>0.03$ and $
1 $ GeV$^2 \ltap Q^2\ltap 6.5$ GeV$^2$,
 while the CERN data extend to $x=0.003$ 
and $1 $ GeV$^2 \ltap Q^2<\ltap 60$ GeV$^2$. The experiments actually 
measure the asymmetry in scattering  {\it i.e. }\  the ratio of the
quantity $a$ of Eqn.(\ref{asym}) to the unpolarized rate of 
Eqn.(\ref{eN}). No $Q^2$ dependence is observed in this ratio. 

In order to extract $G_1(x,Q^2)$ it is assumed that this ratio is
independent of $Q^2$  and therefore that the  $Q^2$ dependence of 
$G_1(x,Q^2)$ is given by that of $F_1(x,Q^2)$ and $F_2(x,Q^2)$.
$G_1(x,Q^2)$ is then extrapolated 
to $Q^2=5$ GeV$^2$. Data 
 are extrapolated to $x=0$ using the assumption that
$G_1(x,Q^2)\sim x^\alpha$ with $0<\alpha <0.5$.
 An extrapolation in the region $0.6 <  x <1$ is 
also needed, but this introduces a
 very small error since $G_1(x,Q^2)$ is very small in this region.
The contribution of the structure function $G_2(x,Q^2)$ 
is suppressed by the $1/Q^2$ term in
Eqn.(\ref{asym}) and no information about it 
can be extracted from the data.

These extrapolations enable the {\sc Spin sum rules}  for $G_1(x,Q^2)$ 
to be evaluated. 
The SMC data alone \cite{SMC93,SMC94,SMC95} give for  
the {\sc Polarized Isospin  sum rule}
\begin{equation}
S_{{\rm Bj}}=0.199\pm0.038
\end{equation}
at $Q^2=10.4$ GeV$^2$, whereas
 at $Q^2$=3 $GeV^2$, the SLAC data \cite{E14293,E14395a,E14395b} give
\begin{equation}
S_{{\rm Bj}}=0.149\pm0.014
\end{equation} 
The different form used for the extrapolation to smaller 
values of $x$ is partly responsible for the smaller values.
The experiments can be combined with earlier results involving  
hydrogen targets \cite{E13083,EMC88} 
to give, at $Q^2=5$ GeV$^2$, \cite{mallot95}
\begin{equation}
\begin{array}{rl}
 S_{{\rm EJ}}^p=&0.136\pm 0.010 \cr
 S_{{\rm EJ}}^n=&-0.067\pm 0.016. 
\end{array}
\end{equation} 
  The {\sc Polarized Isospin  sum rule}  determined from these is
\begin{equation}
S_{{\rm Bj}}=0.203\pm0.023
\end{equation}
\subsection{\it Theory vs. Experiment}
Table 2 shows a comparison of the experimental values discussed above with
theoretical predictions from Table 1. 
In the cases where the experiments have corrected
for the effects of higher twist the
 relevant comparison is with the highest order {\it perturbative} QCD
result available and it is this number
 that is given in the {\it theory} column. No entries are shown for the
$G_2$ and {\sc Second Isospin sum rules} where no data exist.\\
\begin{table}
\begin{tabular}{|c||c|c|c|}
\hline
& $\dsp\frac{Q^2}{{\rm GeV}^2}$  
&  theory 
&  Expt.
\\
\hline
$ S_{{\rm G}}$ 
& 4  &0.335 & 0.216$\pm $0.027\\
\hline
$ S_{{\rm GLS}}$ 
& 3 &2.388 & 2.50 $\pm$ 0.08\\
\hline
$ S_{{\rm A}}$ 
& 3 & 2& 2.02$\pm$ 0.40\\
\hline
$ S_{{\rm Bj}}$ 
& 5 & 0.178  &0.203$\pm$ 0.023
\\
\hline
$ S_{{\rm EJ}}^p$ 
& 5 &0.171  &0.136$\pm$ 0.010
\\
\hline
$ S_{{\rm EJ}}^n$ 
& 5 & -0.0135  &-0.067 $\pm$ 0.016
\\
\hline
\end{tabular} 
\label{tableb}
\caption[]{Comparison of the theoretical
 and experimental values of the sum rules.}
\end{table}
It can be seen from the Table that the sum rules fall
 into three categories. First, the  {\sc Isospin  sum rule}
has very large experimental uncertainties 
but the measured values are consistent with the expectations of QCD.
Second, the {\sc Baryon sum rule} and 
{\sc Polarized Isospin sum rule} are
 compatible with QCD, but have experimental errors that
are small enough so that the measurements can discriminate 
between the QCD results at different orders in perturbation theory.
In these cases the data are consistent with the QCD expectations and
 are inconsistent with the Naive Parton Model.
Finally, the 
{\sc Spin} and {\sc Valence Isospin sum rules} have experimental values
that are inconsistent with the Naive Parton Model or
QCD predictions. 

The second category can be used 
to measure the strong coupling constant $\alpha_s$. Figure 
3 shows the $Q^2$ dependence
of the {\sc Baryon sum rule}. 
This  value is somewhat lower than the world average \cite{hinch95}
The quoted error is dominated by that due to the Higher Twist terms ($HT$). 
The {\sc Polarized Isospin sum rule} has also
 been used to determine $\alpha_s$ \cite{EllKar94}:-
\begin{equation}
\alpha_s(M_Z)=0.122^{+0.005}_{-0.009} 
\end{equation} 
if the higher twist terms are neglected.
or 
\begin{equation}
\alpha_s(M_Z)=0.118^{+0.007}_{-0.014} 
\end{equation} 
if they are included.

The final category needs more discussion. 
The {\sc Valence Isospin sum rule} discrepancy between theory
and experiment shown in table 2 
can be removed by dropping the assumption that
 $\int_0^1dx (\overline{u}(x)-\overline{d}(x))=0$.
Using the Naive Parton Model (see Eqn. \ref{naive-got})
and the data we obtain
\begin{equation}
\int_0^1dx (\overline{u}(x)-\overline{d}(x))=-0.176 \pm 0.040.
\end{equation} 
The QCD corrections are much smaller than the error on this result.
Additional experimental information is available from the processes
 $pd\to \mu^+\mu^-+anything$
and
$pp\to \mu^+\mu^-+anything$ which,
 in the Parton Model, are due to quark  antiquark annihilation.
 Data from NA51 \cite{na51} indicate
that $\overline{u}(x)/\overline{d}(x)=0.51
 \pm 0.08$ at $x=0.18$ and $Q^2=2.5$ GeV$^2$.
The possible non-equality of $\overline{u}(x)$ and $\overline{d}(x)$ was 
first suggested in  \cite{feynman77} where 
 a possible parameterization was introduced.
Several authors have attempted to estimate 
the size of $\overline{u}(x)-\overline{d}(x)$ that could
arise from non-perturbative effects.  Some
 have attempted to explain the effect in terms of a pion cloud
surrounding the nucleon \cite{pion-people}.
  Other models are based a chiral Lagrangian \cite{georgi} approach that
starts with a nucleon consisting of three
 valence quarks then generates the anti-quark distributions from
pions emitted in processes like $u \to d \pi^+$ \cite{quigg91,cheng95}. 
Models of this type produce a difference 
$(\overline{u}(x)-\overline{d}(x))$ 
that is concentrated at very small values of $x$ and the ratio 
$\overline{u}(x)/\overline{d}(x)$ is predicted to be quite small.

The predictions for the values of the {\sc Spin sum rules}
 depend upon the assumed values for
$a_0$, $a_3$ and $a_8$. There is no 
fundamental reason for the first of these to take the value
$a_0=\sqrt{3}a_8$ which was the 
original assumption of \cite{EllJaf74} and is the value used in the 
table.
If we use $a_0=0.12\pm 0.027$, obtained from the value
 of $\Delta S$ obtained from $\nu p$ elastic scattering \cite{kapman88}
we have at $Q^2=5$ GeV$^2$ the ``predictions'' 
$ S_{{\rm EJ}}^n=-0.065\pm 0.030$ and $ S_{{\rm EJ}}^p=0.12\pm 0.03$
both of which agree with the experimental
results shown in Table 2. Instead we can use 
the experimental results for $S_{EJ}^p$ and $S_{EJ}^n$ to determine $a_0$
from the QCD forms of Eqn.(\ref{eljaffcval}) 
together with the higher twist corrections. This 
gives $a_0=0.19\pm 0.07$.
Using the Naive Parton Model relation of Eqn.(\ref{a0eq})
 implies that $\int dx (\Delta s(x)+ \Delta\overline{s}(x))=-0.13\pm 0.
03$ from whence we can also infer that
 $\int dx (\Delta u(x)+ \Delta\overline{u}(x))=0.8$ and 
$\int dx (\Delta d(x)+ \Delta\overline{d}(x))=0.4$.
 Everything is consistent but one is left with the annoying question of what is
carrying most of the nucleon's spin. In the model where 
the nucleon is viewed a soliton-like solution of
\cite{skyrme} one expects $a_0=0$ \cite{Brod88} in the 
limit of zero quark mass and large number of colors. In this interpretation
all of the nucleon's spin is carried by 
 orbital angular momentum. Reference
 \cite{ellis95} can be consulted for a detailed review.

If the gluon distribution in the proton is
 polarized, there is an additional complication. 
The scattering process $e+g\to e q\overline{q} $
 can generate a
contribution to $G_1(x,Q^2)$ at order $\alpha_s$.
 Adding this term to the Naive Parton Model result is
equivalent to replacing Eqn.(\ref{a0eq}) by  \cite{altross}
\begin{equation}
a_0=\int_0^1 dx \Big(\Delta u + \Delta \bar{u}
                +\Delta d + \Delta \bar{d}
                +\Delta s + \Delta \bar{s}
 -n_f\frac{\alpha_s}{2\pi}\Delta g\Big)
\label{aoglue}
\end{equation} 
This contribution is not present in the Operator 
Product analysis presented above. It can be 
introduced if one observes that
while the operator corresponding
 to the singlet axial current is
 not conserved and is therefore subject to renormalization due to
the axial anomaly \cite{jaffe87,carlitz88,efrem88}, a linear 
combination of this operator and a 
 gauge variant operator made up of gluon fields is not 
renormalized. If this term is included the form of the
 QCD corrections given
 in Eqn.(\ref{eljaffcval})
 are the same except that $a_0^{{\rm inv}}$ is 
interpreted as that of 
Eqn.(\ref{aoglue}). The data are now to
be interpreted as implying   $\int dx(\Delta s + 
\Delta \bar{s} -n_f\frac{\alpha_s}{2\pi}\Delta g
 =-0.13 \pm 0.03$. If we assume that
$\Delta S=0$ then  $\int dx \Delta g= 3.2 \pm 0.75$
 at $Q^2=5$ GeV$^2$.  This substantial polarization
 should be observable in
other experiments. For example, the production of
 pions at large transverse  momentum in proton-proton scattering
 proceeds via
parton parton scattering of the type
$q+g \to g q$. If both protons are polarized an asymmetry 
\begin{equation}
A_{LL}=\frac{d\sigma(++)-d\sigma(-+)}{d\sigma(++)+d\sigma(-+)} 
\end{equation} 
can be formed (the $\pm$ arguments refer the helicity of the 
incident protons) which is depends upon 
$\Delta g(z)$. An experiment at FermiLab \cite{adams91} observes
 an asymmetry that is consistent with zero for 
transverse momenta of pions less than 3 GeV. More recently 
\cite{adams94} the same experiment has
measured the asymmetry for double $\pi^0$ production.
 Again the asymmetry is consistent with zero.
If models for $\Delta q$ are assumed \cite{carlitz77}
 then a constraint can be obtained on $\Delta g$.
This constraint is sufficient to rule out some models
 \cite{deadmodels}, but others that have $\int dx \Delta g(x)\sim 5$
\cite{livingmodels} are not excluded.
\section{Conclusions}
The Parton Model
sum rules represent fundamental predictions of QCD.
 The experimental 
precision of many of these rules is such that 
consistency with the theory can 
be established. In the case of a few of the 
rules, notably the {\sc baryon 
sum rule}, the data are sufficiently precise
 that consistency can be
checked in detail and a value of the strong coupling constant obtained
whose error is competitive with the best measurements
 \cite{hinch95}. In this case
the theoretical errors coming from the poor knowledge
 of higher twist terms and the
order $\alpha_s^4$ terms contribute significantly to 
the  error on $\alpha_s$.
Improvement in these areas is unlikely to appear in 
the near future.
The failure of the {\sc Valence Isospin sum rule} has
 led to the realization that
$\overline{u}(x)\neq \overline{d}(x)$ and while there
 is some theoretical understanding
of how this might arise, the difference, like all other
 structure functions, must be extracted from data.
The failure of the naive form of the {\sc Spin sum rules}
 has led to an interesting situation.
There must be significant polarization in the strange
 quarks and/or the gluons. More accurate data
on $\nu p$ elastic scattering might enable the former 
to be constrained. The latter
should be constrained when polarized proton-proton 
scattering experiments become available at RICH in 
the next few years \cite{rich}

The advent of data from HERA\cite{hera} have enabled structure 
functions to be 
measured at smaller values of $x$ than those in fixed target 
experiments. Nevertheless, the statistical errors on these data 
are still quite
large and they are, of course, only available for $F_2^{ep}(x,Q^2)$.
In the future, data from polarized $ep$ scattering will be 
available from this facility \cite{hermes}
that will considerably extend the range of $x$ and $Q^2$ 
available for the
measurements of $G_1(x,Q^2)$ and reduce the error on the
 {\sc Spin
sum rules} resulting from the extrapolation into $x=0$.

\noindent

{\bf Acknowledgments}\\
This work was supported by the Director, Office of
Energy Research, Office of High Energy and Nuclear
 Physics, Division of
High Energy Physics of the U.S. Department of Energy under Contract
DE-AC03-76SF00098 and (AK) by Deutsche Forschungsgemeinschaft
 (DFG) under grant number Kw 8/1-1. Accordingly, the U.S.\
Government retains a nonexclusive, royalty-free license to 
publish or
reproduce the published form of this contribution, or allow others to do
so, for U.S. Government purposes.

\section*{Appendix}
The effective coupling constant in next-next-to-leading
order may be written in the form
\begin{equation}
\ba{ll}\dsp
\frac{\bar{\alpha}_s}{\pi}=
& \dsp
\frac{1}{\beta_0 L}
\Bigg[
1-\frac{1}{\beta_0 L}\frac{\beta_1 \ln L}{\beta_0}
\\ & \dsp
+\frac{1}{\beta_0^2L^2}
\left(
\frac{\beta_1^2}{\beta_0^2}(\ln^2L-\ln L-1)
+\frac{\beta_2}{\beta_0}
\right)
\Bigg]
\ea
\end{equation}
where $L\equiv \ln (Q^2/\Lambda^2)$ and
$\bar{\alpha}_s(Q^2=\mu^2)=\alpha_s(\mu^2)$.
The coefficients of the beta-function are known 
up to the three loop level
\cite{politzer,gross-wil,Cas74,Jon75,TarVlaZha80,LarVer93}:
\begin{equation}
\ba{l}\dsp
\beta_0=\frac{1}{4}\left(11-\frac{2}{3}n_f\right),\;\;\;
\beta_1=\frac{1}{16}\left(102-\frac{38}{3}n_f\right),
\\ \dsp
\beta_2 = \frac{1}{64}\left(
\frac{2857}{2}-\frac{5033}{18}n_f+\frac{325}{54}n_f^2
\right)
\ea
\end{equation}
Anomalous dimensions of singlet and nonsinglet  operators
were calculated in a number of works for both
unpolarized 
\cite{GeoPol74,GroWil74,AltPar77,FloRosSac77,FloRosSac79},
\cite{CurFurPet80,HamNee92,GonLopYnd79,GonLop80,LarTkaVer91a,LarRitVer94} 
and polarized scattering
\cite{AhmRos75,Sas75,AltPar77,Kod80,ZijNee94,MerNee95,Lar93,CheKue93}.
Some of the results are given here.
The coefficients of the nonsinglet anomalous dimension for 
unpolarized scattering read 
\begin{equation}
\ba{l}\dsp
\gamma_{n=1}^{{\rm NS}(0)}
=
-\frac{1}{4}C_F
\left[1-\frac{2}{n(n+1)}+4\sum_{j=2}^n\frac{1}{j}\right]_{n=1}
=  0
\\ \dsp
\gamma_{n=1}^{{\rm NS}(1)+}=-\frac{1}{36}\left(
13+8\zeta(3)-2\pi^2
\right)
\\ \dsp
\gamma_{n=1}^{{\rm NS}(1)-}=0
.\ea
\end{equation}
For polarized scattering one has the following
nonsinglet anomalous dimensions
$\Delta\gamma_{n}^{{\rm NS}(0)}=\gamma_{n}^{{\rm NS}(0)}$
and 
$\Delta\gamma_{n}^{{\rm NS}(1)\pm}=
\gamma_n^{{\rm NS}(1)\pm}$.
Finally the coefficients for the singlet quark diagonal
anomalous dimension matrix elements are
\begin{equation}
\ba{ll}\dsp
\Delta\gamma_{n,qq}^{{\rm S}(0)}
& \dsp = 
\gamma_{n}^{{\rm NS}(0)}
\\  \dsp
\Delta\gamma_{n=1,qq}^{{\rm S}(1)}
=
& \dsp
\frac{1}{16}\left[
\gamma_{n}^{{\rm NS}(1)-}
-4C_Fn_f\frac{n^4+2n^3+2n^2+5n+2}{n^3(1+n)^3}\right]_{n=1} 
\\ & \dsp
=-\frac{1}{2}n_f
\\ \dsp
 \Delta\gamma_{n=1,qq}^{{\rm S}(2)}
=
& \dsp
\frac{1}{64}\left(\left[18 C_F^2-\frac{142}{3}C_FC_A
\right]n_f+\frac{4}{3}C_F n_f^2\right)
\\ & \dsp
= -\frac{59}{24}n_f + \frac{1}{36} n_f^2
\ea
\end{equation} 

Finally we give the analytic formulae of the coefficient
functions corresponding to the first moments
of the various structure functions.
The nonsinglet coefficient function for the structure function
$F_2$ of
unpolarized electron nucleon scattering is given in the 
$\msbar$ scheme 
\cite{BarBurDukMut78,FloRosSac79} by
\begin{equation}
\ba{l}\dsp
 C^{(ep-en)}_{F_2,n=1}(1,\bar{A}(Q^2)) 
=
1+\left(\frac{\alpha_s}{\pi}\right)
\Bigg[ 
\frac{C_f}{4}\Bigg(
-9+\frac{2}{n^2}+\frac{4}{(n+1)}+\frac{3}{n}
\\ \hphantom{xxx}  \dsp
+3\sum_{j=1}^n\frac{1}{j}
-4\sum_{j=1}^n\frac{1}{j^2}
-\frac{2}{n(n+1)}\sum_{j=1}^n\frac{1}{j}
+4\sum_{s=1}^n\frac{1}{s}\sum_{j=1}^s\frac{1}{j}
\Bigg)
\\ \hphantom{xxx} \dsp
-\gamma^{{\rm NS}(0)}_n(\ln(4\pi)-\gamma_E)
\Bigg]_{n=1}
=1
\ea
\end{equation}
The nonsinglet
 \cite{KodMatMutSasUem79,GorLar86,GorLar87,LarVer91,ZijNee94}
and the singlet
 \cite{Kod80,ZijNee94,Lar94}
 coefficient functions for polarized 
electron-nucleon scattering read
\begin{equation}
\ba{l}\dsp
C^{{\rm NS}}_{G_1,n=1}\left(1,\bar{A}(Q^2)\right) 
=  1 - \frac{3}{4}C_F \apib
\\   \hphantom{xxx} \dsp
+ \left( \apib \right)^2 C_F
  \left[
    \frac{21}{32} C_F-\frac{23}{16}C_A+\frac{1}{4}n_f
  \right]
\\  \hphantom{xxx} \dsp
+ \left( \apib \right)^3
  \left[
    -\frac{3}{128} C_F^3
    +C_F^2 C_A \left( \frac{1241}{576}-\frac{11}{12}\zeta(3)
               \right)
  \right.
\\  \hphantom{xxx} \dsp
\hphantom{\left( \apib \right)^3 [}
     +C_F C_A^2\left( -\frac{5437}{864}+\frac{55}{24}\zeta(5)
               \right)
\\  \hphantom{xxx} \dsp
\hphantom{\left( \apib \right)^3 [}
     +C_F^2 n_f\left( -\frac{133}{1152}-\frac{5}{24}\zeta(3)
               \right)
\\  \hphantom{xxx} \dsp
\hphantom{\left( \apib \right)^3 [}
     +C_F C_A n_f\left( \frac{3535}{1728}+\frac{3}{8}\zeta(3)
                        -\frac{5}{12}\zeta(5)
               \right)
\\  \hphantom{xxx} \dsp
\hphantom{\left( \apib \right)^3 [}
\left.
  -\frac{115}{864} C_F n_f^2
\right]\\
\hphantom{xxx} \dsp
= 1-\apib
+\left( \apib \right)^2
    \left[ -\frac{55}{12}+\frac{1}{3}n_f \right]
\\ \hphantom{xxx} \dsp
+\left( \apib \right)^3
\left[
  -\frac{13841}{216}-\frac{44}{9}\zeta(3)+\frac{55}{2}\zeta(5)
\right.
\\ \hphantom{xxx} \dsp
\hphantom{+\left( \apib \right)^3 [}
+ n_f \left(
       \frac{10339}{1296}+\frac{61}{54}\zeta(3)-\frac{5}{3}\zeta(5)
      \right)
\\ \hphantom{xxx} \dsp
\hphantom{+\left( \apib \right)^3 [}
\left.
-\frac{115}{648}n_f^2
\right]
\ea
\end{equation}
\begin{equation}
\ba{l}\dsp
C^{{\rm S}}_{G_1,n=1}\left(1,\bar{A}(Q^2)\right) = 
1-\frac{3}{4}\apib
\\ \hphantom{xxx} \dsp
+ \left(\apib\right)^2\left[
      \frac{21}{32}C_F^2 - \frac{23}{16}C_F C_A
    +C_F n_f\left(
     \frac{13}{48}+\frac{1}{2}\zeta(3)
            \right)       
                     \right] 
\\  \hphantom{xxx} = \dsp
 1 - \apib 
+ \left(\apib\right)^2\left[
             -\frac{55}{12}+n_f\left(
                     \frac{13}{36}+\frac{2}{3}\zeta(3)
                               \right)
                     \right] 
\ea
\end{equation}
The coefficient function for the
  neutrino structure function $F_3$ 
has the following form
 \cite{BarBurDukMut78,LarVer91,ZijNee92b,ZijNee94}:
\begin{equation}
\ba{l}\dsp
C^{(\nu {\cal N})}_{F_3,n=1}
 \left(1,\bar{A}(Q^2)\right) =
  1 - \frac{3}{4}C_F \apib
\\ \hphantom{xxx} \dsp
+ \left( \apib \right)^2 C_F
  \left[
    \frac{21}{32} C_F-\frac{23}{16}C_A+\frac{1}{4}n_f
  \right]
\\ \hphantom{xxx} \dsp
+ \left( \apib \right)^3
  \left[
    -\frac{3}{128} C_F^3
    +C_F^2 C_A \left( \frac{1241}{576}-\frac{11}{12}\zeta(3)
               \right)
  \right.
\\  \hphantom{xxx} \dsp
\hphantom{\left( \apib \right)^3 [}
     +C_F C_A^2\left( -\frac{5437}{864}+\frac{55}{24}\zeta(5)
               \right)
\\  \hphantom{xxx} \dsp
\hphantom{\left( \apib \right)^3 [}
     +C_F^2 n_f\left( -\frac{133}{1152}-\frac{5}{24}\zeta(3)
               \right)
\\  \hphantom{xxx} \dsp
\hphantom{\left( \apib \right)^3 [}
     +C_F C_A n_f\left( \frac{3535}{1728}+\frac{3}{8}\zeta(3)
                        -\frac{5}{12}\zeta(5)
               \right)
\\  \hphantom{xxx} \dsp
\hphantom{\left( \apib \right)^3 [}
  -\frac{115}{864} C_F n_f^2
\\  \hphantom{xxx} \dsp
\hphantom{\left( \apib \right)^3 [}
\left.
  +n_f \frac{d^{abc}d^{abc}}{N_C}
      \left( -\frac{11}{192} + \frac{1}{8}\zeta(3) \right)
\right]
\ea
\end{equation}
\begin{equation}
\ba{ll}\dsp
= 
& \dsp
1-\apib
+\left( \apib \right)^2
    \left[ -\frac{55}{12}+\frac{1}{3}n_f \right]
\\ & \dsp
+\left( \apib \right)^3
\left[
  -\frac{13841}{216}-\frac{44}{9}\zeta(3)+\frac{55}{2}\zeta(5)
\right.
\\ & \dsp
\hphantom{+\left( \apib \right)^3 [}
+ n_f \left(
       \frac{10009}{1296}+\frac{91}{54}\zeta(3)-\frac{5}{3}\zeta(5)
      \right)
\\ & \dsp
\hphantom{+\left( \apib \right)^3 [}
\left.
-\frac{115}{648}n_f^2
\right]
\ea
\end{equation}
The coefficient function for the 
nonsinglet structure function $F_1$ of neutrino
and antineutrino scattering reads
\cite{BarBurDukMut78,CheGorLarTka84,LarTkaVer91}
\begin{equation}
\ba{l}\dsp
 C^{(\nu n-\nu p)}_{F_1,n=1}(1,\bar{A}(Q^2)) 
=
  1 - \frac{1}{2}C_F \apib
\\  \hphantom{xxx} \dsp
+ \left( \apib \right)^2 C_F
  \left[
    \frac{11}{16} C_F-\frac{91}{72}C_A+\frac{2}{9}n_f
  \right]
\\ \hphantom{xxx} \dsp
+ \left( \apib \right)^3
  \left[
     C_F^3 \left(
         -\frac{313}{64}-\frac{47}{4}\zeta(3)+\frac{35}{2}\zeta(5)
           \right)
  \right.
\\ \hphantom{xxx} \dsp
\hphantom{\left( \apib \right)^3 [}
    +C_F^2 C_A \left( 
        \frac{2731}{288}+\frac{91}{6}\zeta(3)-\frac{95}{4}\zeta(5)
               \right)
\\ \hphantom{xxx} \dsp
\hphantom{\left( \apib \right)^3 [}
     +C_F C_A^2\left( 
        -\frac{8285}{1296}-\frac{5}{2}\zeta(3)+5\zeta(5)
               \right)
\\ \hphantom{xxx} \dsp
\hphantom{\left( \apib \right)^3 [}
     +C_F^2 n_f\left( -\frac{335}{576}+\frac{1}{12}\zeta(3)
               \right)
\\ \hphantom{xxx} \dsp
\hphantom{\left( \apib \right)^3 [}
     +C_F C_A n_f\left( \frac{4235}{2592}-\frac{7}{12}\zeta(3)
                        +\frac{5}{6}\zeta(5)
               \right)
\\ \hphantom{xxx} \dsp
\hphantom{\left( \apib \right)^3 [}
\left.
  -\frac{155}{1296} C_F n_f^2
\right]
\ea
\end{equation}
\begin{equation}
\ba{ll}\dsp
= 
& \dsp
1-\frac{2}{3}\apib
+\left( \apib \right)^2
    \left[ -\frac{23}{6}+\frac{8}{27}n_f \right]
\\ & \dsp
+\left( \apib \right)^2
\left[
  -\frac{4075}{108}+\frac{622}{27}\zeta(3)-\frac{680}{27}\zeta(5)
\right.
\\ & \dsp
+ n_f \left(
       \frac{3565}{648}-\frac{59}{27}\zeta(3)+\frac{10}{3}\zeta(5)
      \right)
\left.
-\frac{155}{972}n_f^2
\right]
\ea
\end{equation}

\end{document}